\documentclass{aastex63}

\usepackage{graphicx}
\usepackage{amsmath}

\usepackage{epsfig}

\usepackage{subfigure}
\usepackage{longtable}
\usepackage{times}

\shorttitle{Comparison of the Kerr and Einstein-Gauss-Bonnet gravities}
\shortauthors{O.Donmez, F. Dogan and T. Sahin}

\begin{document}

\title{Study of Asymptotic Velocity in the Bondi-Hoyle Accretion Flows in the Domain of
  Kerr and 4-D Einstein-Gauss-Bonnet Gravities}
      
\author{Orhan Donmez}
\altaffiliation{College of Engineering and Technology, American
  University of the Middle East, Kuwait}

\author{Fatih Dogan}
\altaffiliation{College of Engineering and Technology, American
  University of the Middle East, Kuwait}

\author{Tuba Sahin}
\altaffiliation{College of Engineering and Technology, American
  University of the Middle East, Kuwait}

\begin{abstract}
  Understanding the physical structures of the accreated matter very close to the black hole in
  quasars and active galactic nucleus (AGNs) is an important milestone to constrain the activities occurring
  in their centers. In this paper, we numerically investigate the effects of the asymptotic velocities
  on the physical structures of the accretion disk around the Kerr and Einstein-Gauss-Bonnet (EGB) rapidly rotating black holes.
  The Bondi-Hoyle accretion is considered with a falling gas towards the black hole in upstream region of the computational domain.
  The shock cones are naturally produced in the downstream part of the flow around both black holes.
  It is found that the structure of the cones and the amount of the accreated matter depend on
  asymptotic velocity $V_{\infty}$ (Mach number) and the types of the gravities (Kerr or EGB).
  Increasing the Mach number of the inflowing matter in the supersonic region causes the shock opening angle and
  accretion rates getting smaller because of  the rapidly falling gas towards the black hole.
  The EGB gravity leads to  an increase in the shock opening angle
  of the shock cones while the mass accretion rates $\dot{M}$ are decreasing in EGB gravity with 
  a Gauss-Bonnet (GB) coupling constant $\alpha$. It is also confirmed that accretion rates  and drag forces
  are significantly altered in the EGB gravity. Our numerical simulation results could be used
  to identify the accreation mechanism and physical properties of the accretion disk  and black hole in the
  observed $X-$ rays such as NGC $1313$ $X-1$ and $1313$ $X-2$ and MAXI $J1803-298$.

\end{abstract}


%%%%%%%%%%%%%%%%%%%%%%%%%%%%%%%%%%%%%%%%%%%%%%%%%%%%%%%%%%%%%%%%%%%%%%%
%%%%%%%%%%%%%%%%%%%%%%%%%%%%%%%%%%%%%%%%%%%%%%%%%%%%%%%%%%%%%%%%%%%%%%%
%%%%%%%%%%%%%%%%%%%%%%%%%%%%%%%%%%%%%%%%%%%%%%%%%%%%%%%%%%%%%%%%%%%%%%%
%%%%%%%%%%%%%%%%%%%%%%%%%%%%%%%%%%%%%%%%%%%%%%%%%%%%%%%%%%%%%%%%%%%%%%%
%%%%%%%%%%%%%%%%%%%%%%%%%%%%%%%%%%%%%%%%%%%%%%%%%%%%%%%%%%%%%%%%%%%%%%%
%%%%%%%%%%%%%%%%%%%%%%%%%%%%%%%%%%%%%%%%%%%%%%%%%%%%%%%%%%%%%%%%%%%%%%%

\section{Introduction}
\label{Introduction}

The theory of the general relativity has already been confirmed by the large number of observational
tests such as gravitational wave by LIGO \citep{LIGO1}, the $M87*$ black hole shadow by
Event Horizon Telescope (EHT) \citep{Khodadi1,Allahyari1,Akiyama1,Akiyama2,Akiyama3},
and the emitted electromagnetic spectrum  from an accreation disk \citep{Yuan1}. In the
strong gravitational region, the dynamics of the accretion disk close to the black hole uncovers
the properties of the black hole and electromagnetic spectrum. It is certain that this region is
influenced by the space-time curvature which depends on the black hole spin $a$ and 
GB coupling constant $\alpha$ in General Relativity (GR) and $4D$
Einstein-Gauss-Bonnet (EGB) gravity \citep{Glavan1}.

The wind accretion onto the black hole potentially produces energetic outflows which can be used to define
the black hole spin, mass and its shadow. The magnetically driven outflow onto the black hole
is called Bondi-Hoyle-Lyttleton (BHL) accretion which is purely hydrodynamical \citep{Hoyle1, Bondi1, Edgar1}.
The outflowing gas causes a formation of shock cone due to strong gravity in the vicinity of the black hole.
For decades, BHL accretion was studied using the Newtonian and general relativistic hydrodynamics.
Newtonian hydrodynamic was used to define the structure of the disk due to axisymmetric accretion flow
for adiabatic gas \citep{Hunt1}. $2D$ and $3D$ numerical simulations of the accretion disk had accomplished
to reveal the dynamical structure of the disk, radiation mechanism, and properties of the compact objects
\citep{Foglizzo1, Blondin1, MacLeod1, Ohsugi1, Wenrui1}.
Non-magnetized or magnetized relativistic BHL accretions around
the non-rotating and rotating black holes have been simulated using either axial or spherical symmetries
\citep{Donmez6, Penner1, Donmez5, Penner2, LoraClavijo1,  Koyuncu1, LoraClavijo2, Gracia-Linares1,
  CruzOsorio1}.

The modified theory of gravity received lots of attention when considering the solution of accretion disks.
The accretion disk properties and their dynamical evolutions were studied in different modified gravity
models
such as the innermost circular orbits of spinning and massive particles \citep{Zhang1,Minyong1},
$f(R)$ gravity \citep{Pun1, Staykov1}, Einstein-Maxwell-dilation theory  \citep{Karimow1, Haydari1},
scalar-tensor-vector gravity \citep{Perez1}, Einstein-scalar-Gauss-Bonnet gravity \citep{Haydari2},
$4D$ $EGB$ gravity for non-rotating black hole \citep{Liu1}, the observational constrain the
$GB$ coupling constant $\alpha$ \citep{Feng1,Timothy1},
and for rotating black hole \citep{Fard1}.
\citet{Fard1} had studied the thin accretion disk around rapidly rotating black hole. It is believed that
the black hole would rotate with high rotation velocity due to the accreation effect.

The study of the dynamical evolution of the accretion disk around the non-rotating and
the rotating black holes using different gravities would allow us to extract detailed
information about the central objects, such as black hole shadow and physical properties as
well as emission spectrum and temperature distribution of the accreation disk. The analytic
solutions of the thin accreation disk around the $4D$ EGB
gravity were studied in
\citet{Gyulchev1, Fard1, Guo1, Liu1, Malafarina1} and referenced therein.
They defined the electromagnetic properties  of the
disk and investigated the effects of GB coupling constant $\alpha$ and black hole
rotation parameter $a$ on the properties of the accretion disk, the energy flux,
and the electromagnetic spectrum. They also compared
their results with the Kerr black hole solution. According to their results, the disk around the $4D$ EGB black hole
for the positive value of $\alpha$ is more luminous and hotter than the one in
General Relativity GR \citep{Fard1}. The numerical investigation of a BHL accretion
in the $4D$ EGB extensively was studied in the vicinities of the non-rotating  \citep{Donmez3}
and the rotating black holes \citep{Donmez_EGB_Rot}. They discussed the effect of
GB coupling constant $\alpha$ on the shock cone structure created during the formation  of the
accretion disk.

The aim of this work is to model the accretion disk dynamics in the presence of
the $4D$ EGB and Kerr strong gravities  and to compare the shock cone properties from these two gravities.
The matter will be accreated with a mechanism called
a BHL accretion. The traveling black hole through a uniform medium causes
the BHL accretion and the accreated matter towards the black hole
from upstream side of computational domain forms a steady-state disk around the black hole.
Since we are interested in the dynamics of the accretion disk, 
the shock cone, and accretion efficiency by  using the different gravities,
we explicitly study the effect of  GB coupling constant $\alpha$
and the black hole rotation parameter $a$ on these dynamics. 

In this paper, we model the non-magnetized  BHL accretion onto the spinning black holes in
$4D$ EGB and Kerr strong gravity regions to have a direct comparison between the two gravities.
In section \ref{GRHE1}, brief descriptions of $4D$ EGB and Kerr rotating black hole space-time
metric are presented along with the general relativistic hydrodynamical equations. The initial and
boundary conditions used in numerical simulations are given in Section \ref{GRHE2} in order to inject
the gas from outer boundary of the computational domain. In Section \ref{Results}, we present the
results of our numerical simulations and discuss the consequences of two different gravities on the disk and
shock cone dynamics. In Section \ref{AstroMot}, the astrophysical motivation of the numerical results  is
briefly discussed. The discussion and summary are presented in Section \ref{Conclusion}.

%%%%%%%%%%%%%%%%%%%%%%%%%%%%%%%%%%%%%%%%%%%%%%%%%%%%%%%%%%%%%%%%%%%%%%%
%%%%%%%%%%%%%%%%%%%%%%%%%%%%%%%%%%%%%%%%%%%%%%%%%%%%%%%%%%%%%%%%%%%%%%%

\section{Rotating Black Hole Metric and General Relativistic Equations}
\label{GRHE1}

The Bondi-Hoyle accretion of the perfect fluid in the case of rotating -
Kerr and EGB black holes  is studied by solving General Relativistic
Hydrodynamical (GRH) equations in the curved background. The perfect fluid
stress-energy-momentum tensor is

\begin{eqnarray}
 T^{ab} = \rho h u^{a}u^{b} + P g^{ab},
\label{GREq1}
\end{eqnarray}

\noindent $h$, $p$, $\rho$, $u^{a}$, and $g^{ab}$ are the specific enthalpy,
the fluid pressure, the rest-mass density, the $4-$ velocity of the fluid, and
the metric of the curved space-time, respectively. The indexes $a$, $b$ and $c$ go from $0$ to $3$.
Two different coordinates are used
to compare the dynamical evolution of the accretion disk around the rotating black
hole. First, Kerr black hole in Boyer-Lindquist coordinate is \citep{Donmez6}

\begin{eqnarray}
  ds^2 = -\left(1-\frac{2Mr}{\sum^2}\right)dt^2 - \frac{4Mra}{\sum^2}sin^2\theta dt d\phi
  + \frac{\sum^2}{\Delta_1}dr^2 + \sum^2 d\theta^2 + \frac{A}{\sum^2}sin^2\theta d\phi^2
\label{GREq2}
\end{eqnarray}

\noindent where $\Delta_1 = r^2 - 2Mr +a^2$, and
$A = (r^2 + a^2)^2 - a^2\Delta sin^2\theta$.
The lapse function and shift vector of the Kerr metric are
$\tilde{\alpha} = (\sum^2 \Delta_1/A)^{1/2} $ and $\beta^i = (0,0,-2Mar/A)$.

\noindent Second, the rotating black hole metric in $4D$ EGB gravity is \citep{Donmez_EGB_Rot}

\begin{eqnarray}
  ds^2 &=& -\frac{\Delta_2 - a^2sin^2\theta}{\Sigma}dt^2 + \frac{\Sigma}{\Delta_2}dr^2 -
  2asin^2\theta\left(1- \frac{\Delta_2 - a^2sin^2\theta}{\Sigma}\right)dtd\phi + 
  \Sigma d\theta^2 + \nonumber \\
  && sin^2\theta\left[\Sigma +  a^2sin^2\theta \left(2- \frac{\Delta_2 -  
   a^2sin^2\theta}{\Sigma} \right)  \right]d\phi^2,
\label{GREq3}
\end{eqnarray}

\noindent
where $\Sigma$ and $\Delta_2$ read as $\Sigma = r^2 + a^2cos^2\theta$ and
$\Delta_2 = r^2 + a^2 + \frac{r^4}{2\alpha}\left(1 - \sqrt{1 + \frac{8 \alpha M}{r^3}} \right)$.
$a$, $\alpha$, and $M$ are spin parameter, Gauss-Bonnet coupling constant, and
mass of the black hole, respectively. The horizons of the black holes were obtained
by solving $\Delta_1=0$  and $\Delta_2=0$.  The lapse function $\tilde{\alpha}$ and the shift vectors
of the EGB metric are
$\tilde{\alpha} = \sqrt{\frac{a^2(1-f(r))^2}{r^2+a^2(2-f(r))} + f(r)}$ and $\beta^i =
(0,\frac{a r^2}{2\pi \alpha}\left(1 -\sqrt{1 + \frac{8 \pi \alpha M}{r^3}}\right),0)$,
respectively.  $f(r) = 1 + \frac{r^2}{2\alpha}\left(1 -
\sqrt{1 + \frac{8 \alpha M}{r^3}} \right).$

\noindent In order to solve GRH equation numerically, we should write them in a conserved form
\citep{Donmez1}:

\begin{eqnarray}
  \frac{\partial U}{\partial t} + \frac{\partial F^r}{\partial r} + \frac{\partial F^{\phi}}{\partial \phi}
  = S,
\label{GREq4}
\end{eqnarray}

\noindent where $U$, $F^r$, $F^{\phi}$, and $S$ are the vectors of the conserved variables, of the fluxes along
$r$ and $\phi$ directions, sources, respectively. The vectors of the conserved variables are written in terms
of the primitive variables,

\begin{eqnarray}
  U =
  \begin{pmatrix}
    D \\
    S_j \\
    \tau
  \end{pmatrix}
  =
  \begin{pmatrix}
    \sqrt{\gamma}W\rho \\
    \sqrt{\gamma}h\rho W^2 v_j\\
    \sqrt{\gamma}(h\rho W^2 - P - W \rho)
    \end{pmatrix}
\label{GREq5}
\end{eqnarray}

\noindent where $W = (1 - \gamma_{a,b}v^i v^j)^{1/2}$,  $h = 1 + \epsilon + P/\rho$, $\epsilon$, and
$v^i = u^i/W + \beta^i$ are the Lorentz factor, enthalpy, internal energy, and three-velocity of the fluid,
respectively. The ideal gas equation of state is adopted to define the pressure of the fluid and
three-metric $\gamma_{i,j}$ and its determinant $\gamma$ are computed using the four-metric of
rotating black holes. Latin indices $i$ and $j$ go from $1$ to $3$. The flux and the source can be computed
for any metric using the following equations,

\begin{eqnarray}
  \vec{F}^i =
  \begin{pmatrix}
    \tilde{\alpha}\left(v^i - \frac{1}{\tilde{\alpha}\beta^i}\right)D \\
    \tilde{\alpha}\left(\left(v^i - \frac{1}{\tilde{\alpha}\beta^i}\right)S_j + \sqrt{\gamma}P\delta^i_j\right)\\
    \tilde{\alpha}\left(\left(v^i - \frac{1}{\tilde{\alpha}\beta^i}\right)\tau  + \sqrt{\gamma}P v^i\right)
    \end{pmatrix}
\label{GREq6}
\end{eqnarray}

\noindent and,

\begin{eqnarray}
  \vec{S} =
  \begin{pmatrix}
    0 \\
    \tilde{\alpha}\sqrt{\gamma}T^{ab}g_{bc}\Gamma^c_{aj} \\
    \tilde{\alpha}\sqrt{\gamma}\left(T^{a0}\partial_{a}\tilde{\alpha} - \tilde{\alpha}T^{ab}\Gamma^0_{ab}\right)
   \end{pmatrix} 
\label{GREq7}
\end{eqnarray}

\noindent where $\Gamma^c_{ab}$ is the Christoffel symbol.

%%%%%%%%%%%%%%%%%%%%%%%%%%%%%%%%%%%%%%%%%%%%%%%%%%%%%%%%%%%%%%%%%%%%%%%
%%%%%%%%%%%%%%%%%%%%%%%%%%%%%%%%%%%%%%%%%%%%%%%%%%%%%%%%%%%%%%%%%%%%%%%
\section{Initial and Boundary Conditions}
\label{GRHE2}

To study the Bondi-Hoyle accretion onto the rotating Gauss-Bonnet black hole and compare
it with the Kerr black hole, GRH equations are solved on the equatorial plane using the code
explained in \citet{Donmez1, Donmez5, Donmez2}. The pressure of the accreated matter is handled
by using the standard $\Gamma$ law equation of state for a perfect fluid,
$P = (\Gamma - 1)\rho\epsilon$ with $\Gamma = 4/3$. We adjusted the initial density and pressure
profiles to ensure that the speed of sound  equals to $C_{\infty}=0.1$ at the upstream region of the
computational domain. After setting the density to be a constant value $(i.e.\; \rho =1)$, 
the pressure is computed from the perfect fluid equation of state, then we perform the numerical
simulation on the equatorial plane using
different values of asymptotic velocities. The initial velocities  of the falling matter are given
in terms of an asymptotic velocity $V_{\infty}$ at the upstream region and they are  defined by the following equations,
\citep{Donmez6,Donmez5, Donmez3}

\begin{eqnarray}
  v^r&=& \sqrt{\gamma^{rr}}V_{\infty}cos(\phi) \nonumber \\
  v^{\phi}&=& -\sqrt{\gamma^{\phi \phi}}V_{\infty}sin(\phi).
\label{GRH1}
\end{eqnarray}

\noindent Using various values of $V_{\infty}$ allows us to investigate  different regimes
around the black hole and therefore consider subsonic and supersonic accretion. The matter injected in
the upstream region falls into  a region assumed empty with $V^r=0$, $V^{\phi}=0$, and
$\rho =10^{-4}$ in the beginning of the simulation $(i.e.\; t=0)$. The description of the initial models around
the rotating black holes, Kerr and Gauss-Bonnet, with a spin $a/M=0.97$ is reported in
Table \ref{Inital_Con}.

\begin{table}
%\scriptsize
\footnotesize
\caption{Adopted initial model used in the numerical simulation for Kerr and Gauss-Bonnet black holes.
  $type$ is the black hole types used in numerical simulations.
  $\alpha$ is Gauss-Bonnet coupling constant,
  $V_{\infty}$ is the asymptotic velocity of the matter falling towards the black hole at infinity,
 $M_{\infty} = V_{\infty}/C_{\infty}$ is the asymptotic Mach number at infinity.}
 \label{Inital_Con}
\begin{center}
%\vspace*{2cm}
  \begin{tabular}{ccccc}
    \hline
    \hline

    $type$ & $\alpha (M^2)$ & $V_{\infty}$ & $M_{\infty}$\\
    \hline
    & $-$   & $0.1$ & $1$\\
    & $-$   & $0.2$ & $2$\\
    $Kerr$  & $-$  & $0.3$ & $3$\\
    & $-$   & $0.4$ & $4$\\
    & $-$   & $0.5$ & $5$\\
    & $-$   & $0.6$ & $6$\\
    \hline
    & $0.0064$   & $0.1$ & $1$\\
    & $0.0064$   & $0.2$ & $2$\\
    $Gauss-Bonnet$ & $0.0064$   & $0.3$ & $3$\\
    & $0.0064$  & $0.4$ & $4$\\
    & $0.0064$   & $0.5$ & $5$\\
    & $0.0064$   & $0.6$ & $6$\\
    \hline
    & $-0.1088$   & $0.1$ & $1$\\
    & $-0.1088$   & $0.2$ & $2$\\
    $Gauss-Bonnet$ & $-0.1088$   & $0.3$ & $3$\\
    & $-0.1088$  & $0.4$ & $4$\\
    & $-0.1088$   & $0.5$ & $5$\\
    & $-0.1088$   & $0.6$ & $6$\\
    \hline
    & $-0.5912$   & $0.01$ & $0.1$\\
    & $-0.5912$   & $0.05$ & $0.5$\\    
    & $-0.5912$   & $0.1$ & $1$\\
    & $-0.5912$   & $0.2$ & $2$\\
    $Gauss-Bonnet$ & $-0.5912$   & $0.3$ & $3$\\
    & $-0.5912$  & $0.4$ & $4$\\
    & $-0.5912$   & $0.5$ & $5$\\
    & $-0.5912$   & $0.6$ & $6$\\       
    \hline
    \hline
  \end{tabular}
\end{center}
%  \tablenotetext{}{}
%\vskip -0.8truecm
\end{table}

The uniformly spaced zones are used in radial and angular directions $N_r=1024 $ and $N_{\phi}=256$.
The inner and outer boundaries of the computational domain in the radial direction
are located at $r_{min}=2.3M$ and $r_{max}=100M$,
respectively, and $\phi_{min}=0$ and $\phi_{max}=2\pi $ in the angular direction. It is confirmed that
the qualitative results of the numerical solutions (i.e. appearance of QPOs and instabilities, the shock
location, the behavior of the accretion rates) are not sensitive to the grid resolution.

The corrected treatment of the boundary is important to avoid unphysical solutions in the
numerical simulations. At the inner radial boundary, we implement outflow boundary condition to let
the gas falling into a black hole by a simple zeroth-order extrapolation. On the other hand, we have to
distinguish the downstream $-\pi/2 < \phi < \pi/2$ and the upstream region $\pi/2 \leq \phi \leq 3\pi/2$.
While we adopt the outflow boundary condition in the downstream region, the gas is injected continuously  with
the initial velocities given in Eq.\ref{GRH1} in the upstream region. The periodic boundary is used along the 
$\phi$-direction.

%%%%%%%%%%%%%%%%%%%%%%%%%%%%%%%%%%%%%%%%%%%%%%%%%%%%%%%%%%%%%%%%%%%%%%%
%%%%%%%%%%%%%%%%%%%%%%%%%%%%%%%%%%%%%%%%%%%%%%%%%%%%%%%%%%%%%%%%%%%%%%%
\section{The Numerical Simulation of the accretion onto 4D EGB Rotating and Kerr Black Holes}
\label{Results}

%%%%%%%%%%%%%%%%%%%%%%%%%%%%%%%%%%%%%%%%%%%%%%%%%%%%%%%%%%%%%%%%%%%%%%

\begin{figure*}
  \vspace{1cm}
  \center
  \psfig{file=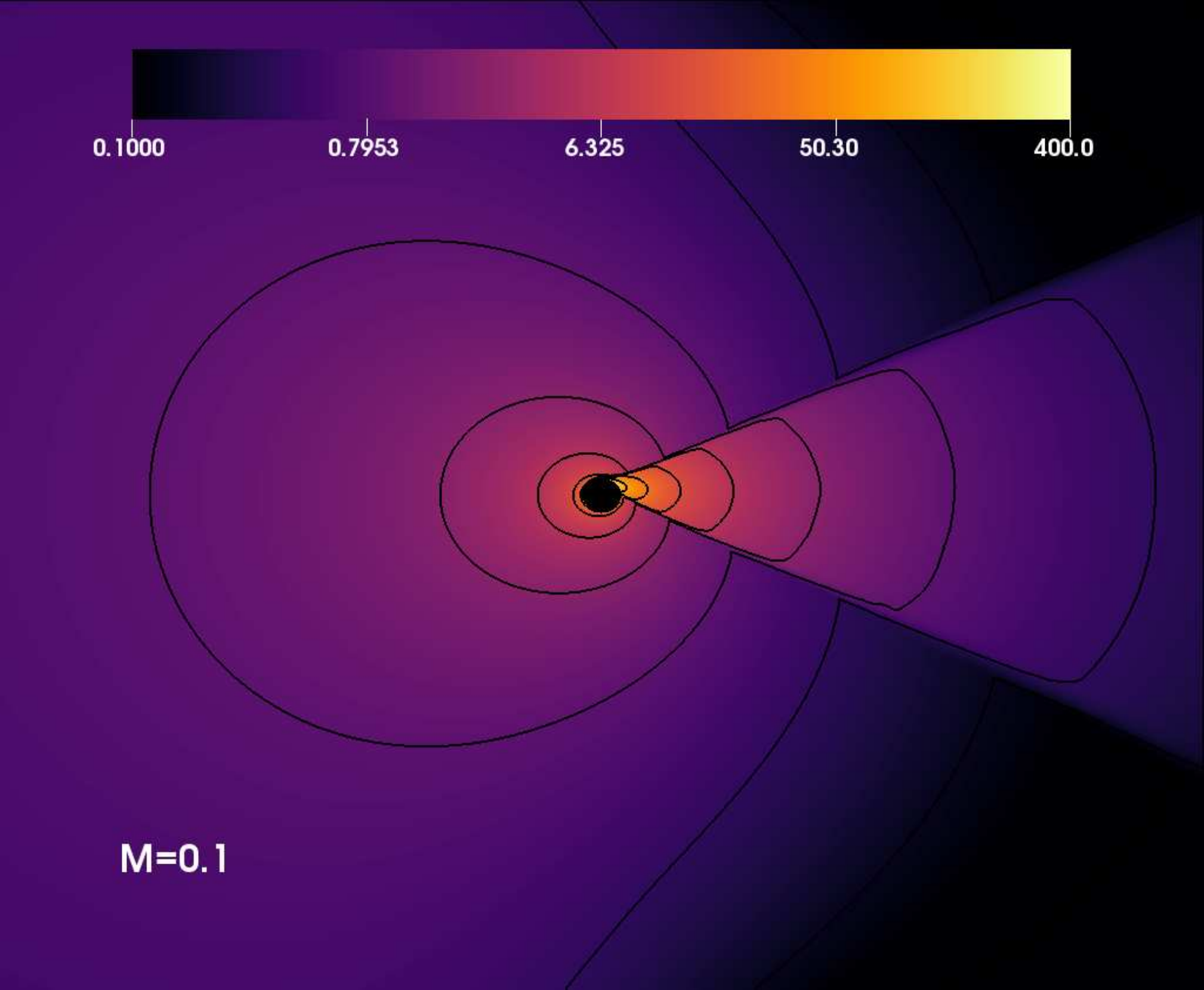,width=6.5cm}
  \psfig{file=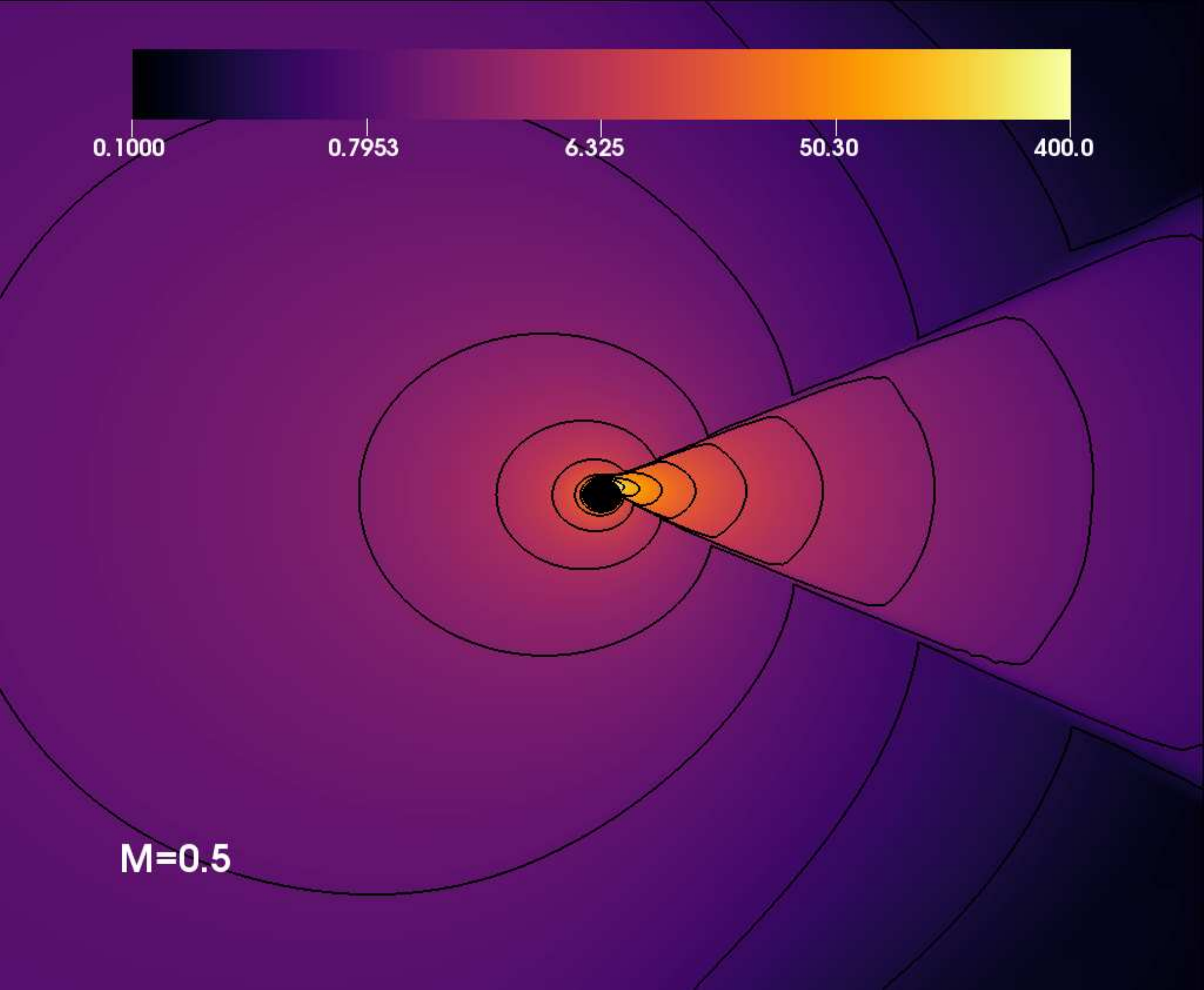,width=6.5cm}
  \psfig{file=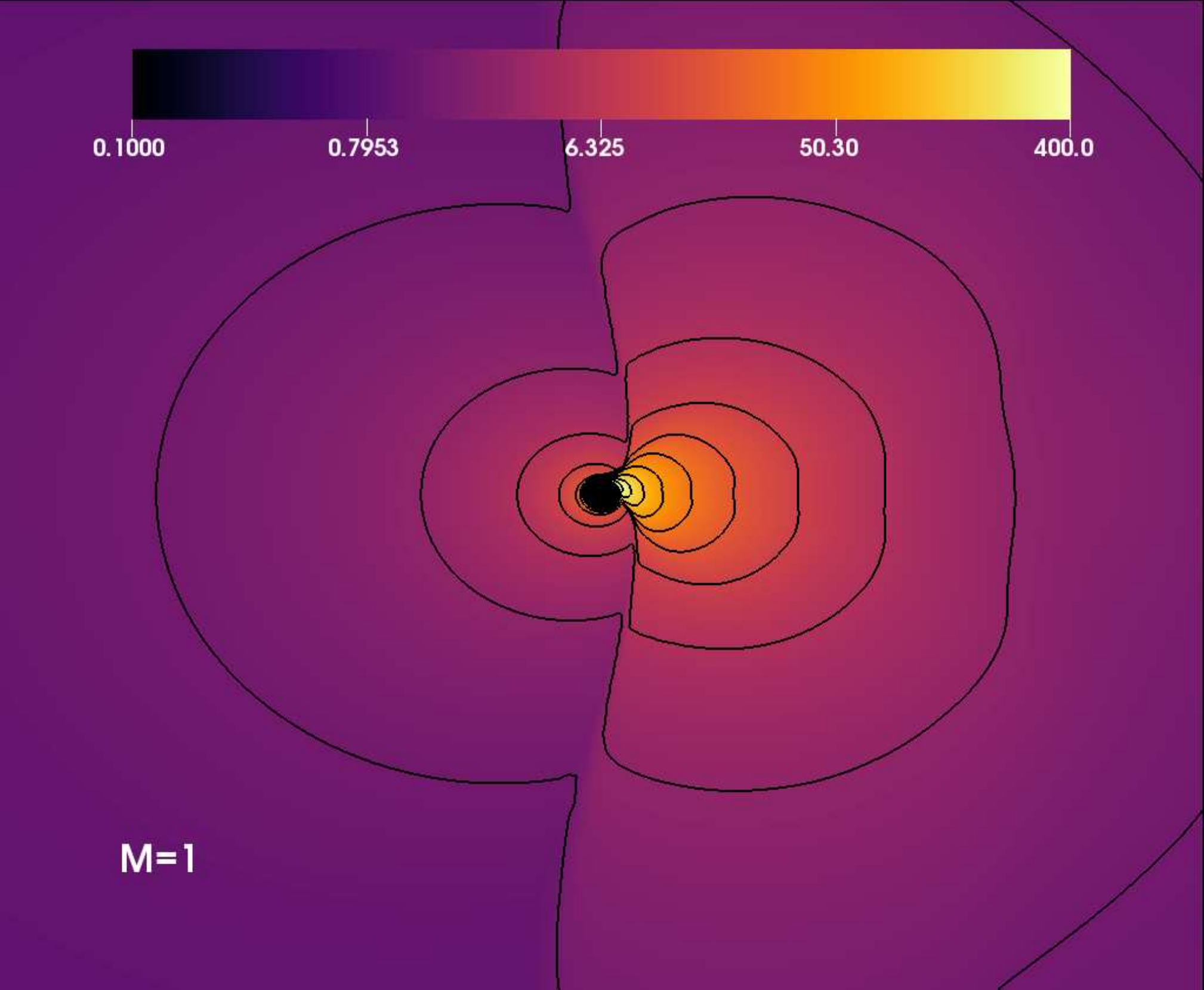,width=6.5cm}
  \psfig{file=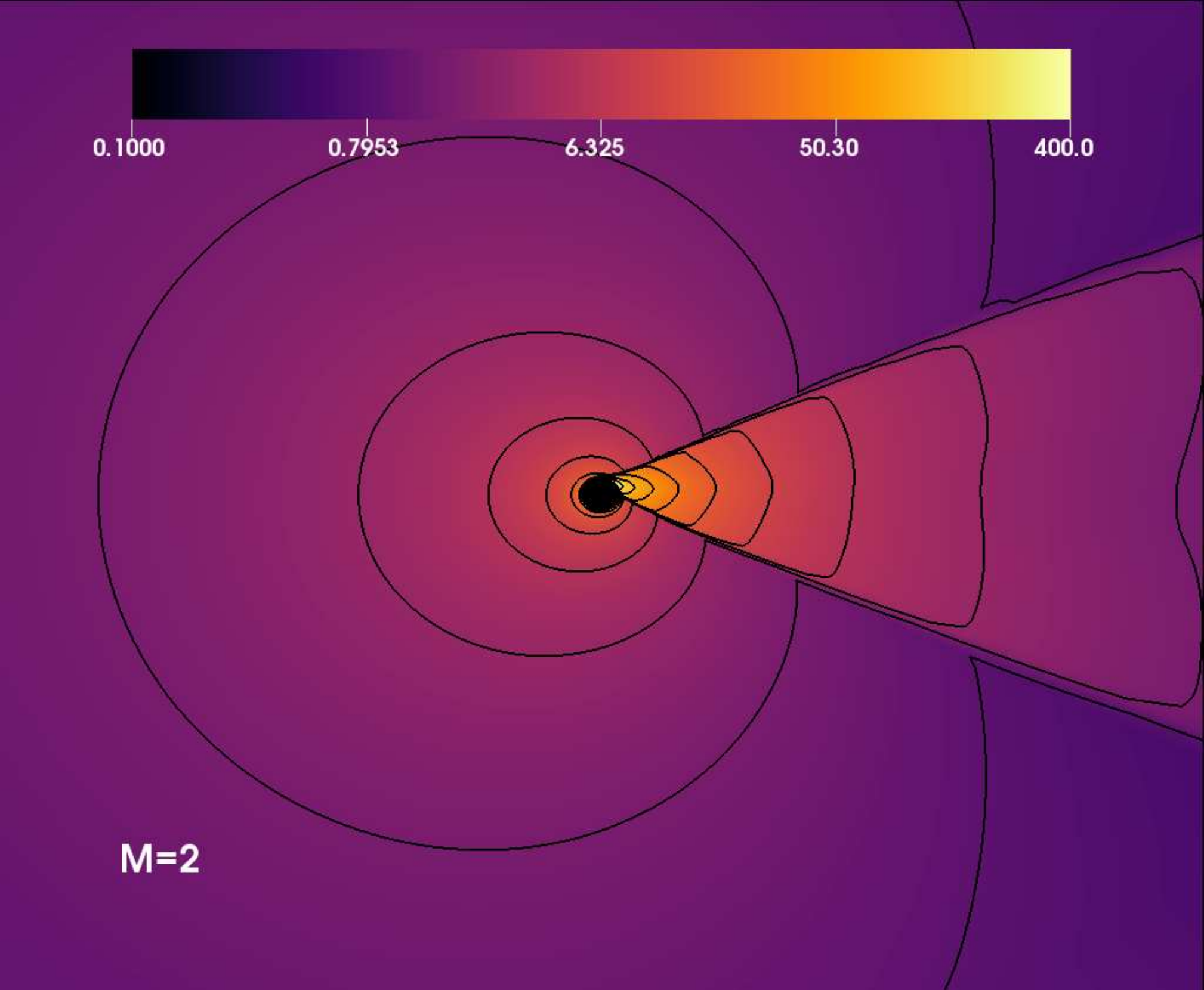,width=6.5cm}
  \psfig{file=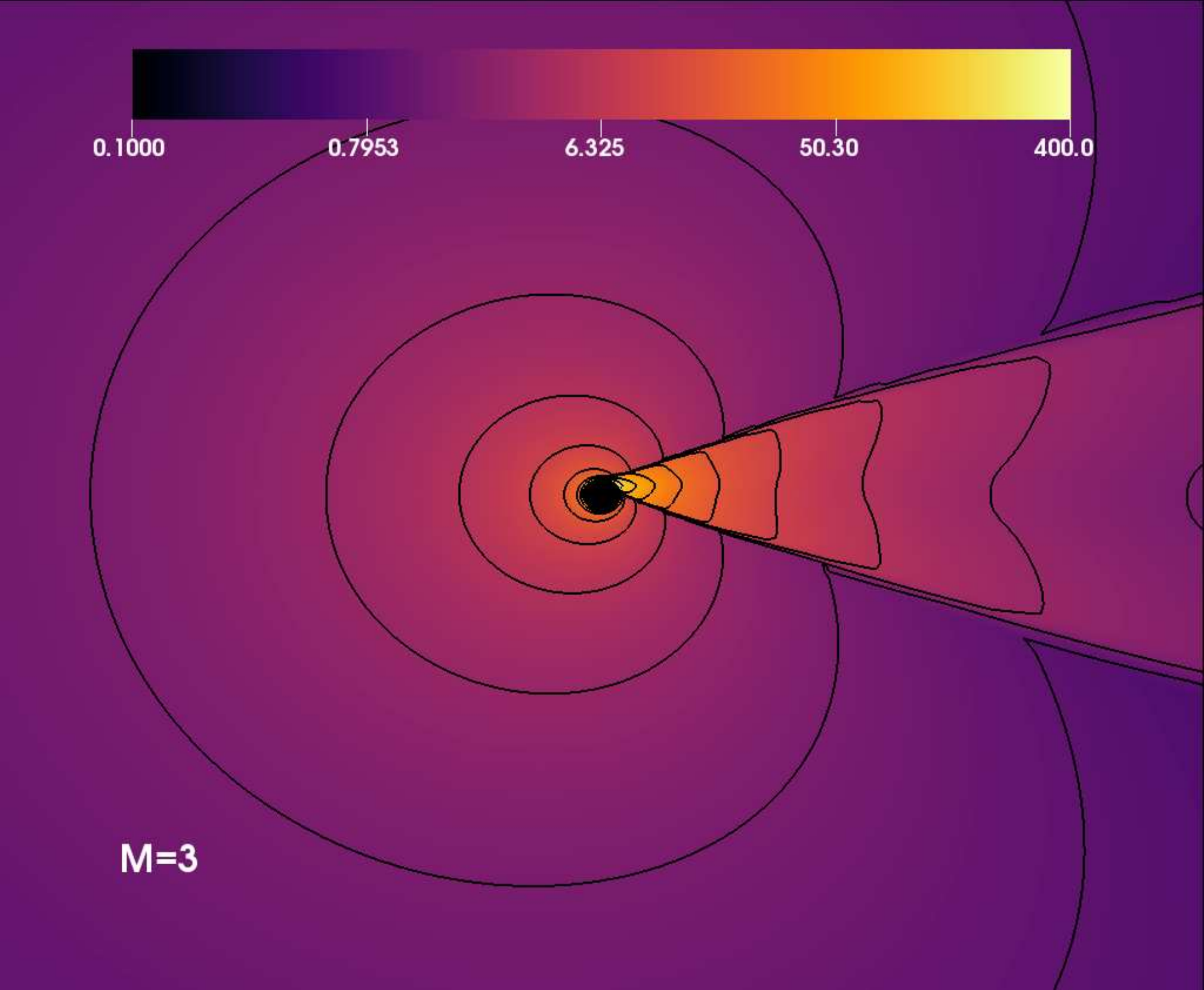,width=6.5cm}
  \psfig{file=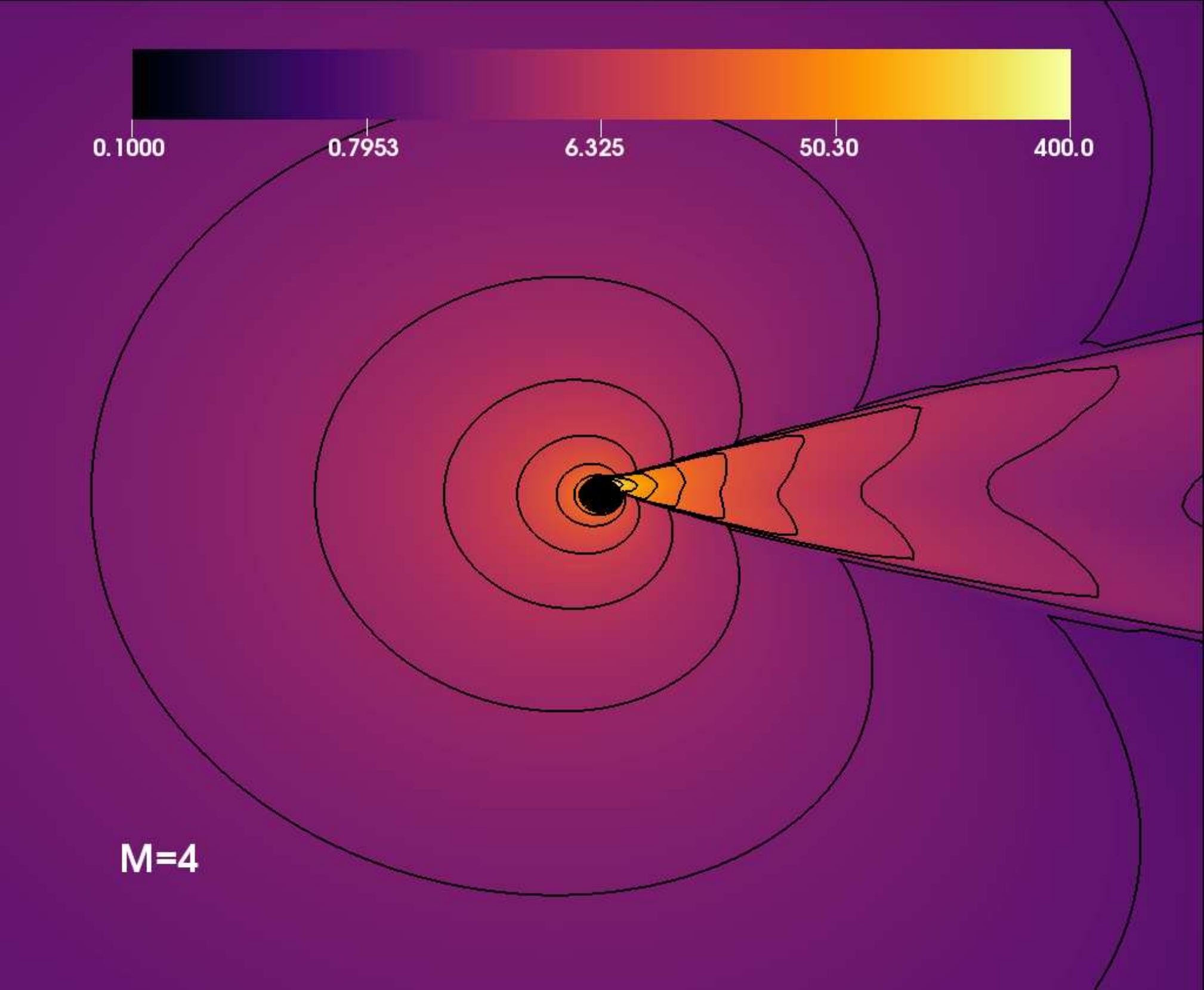,width=6.5cm}
  \psfig{file=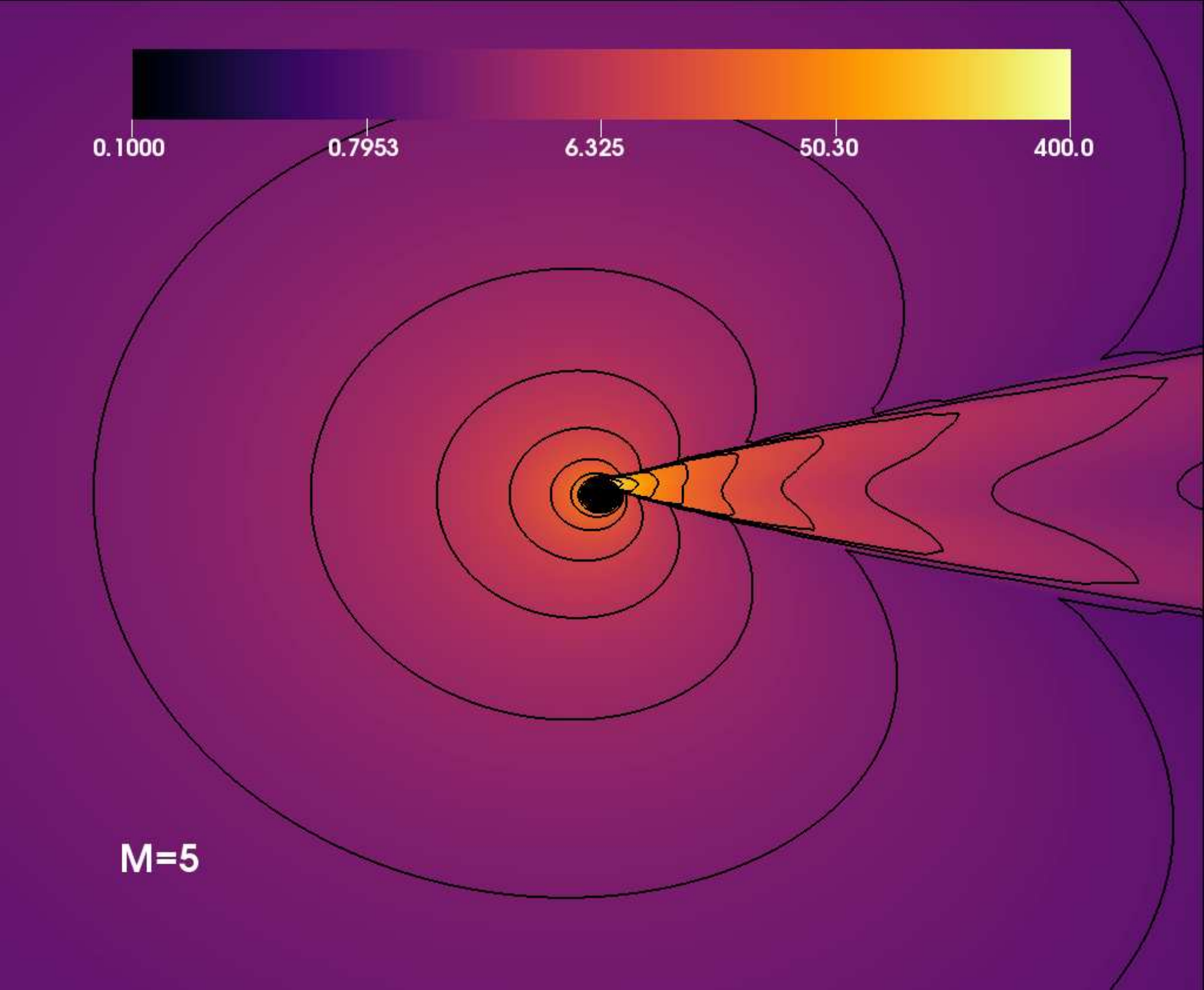,width=6.5cm}
  \psfig{file=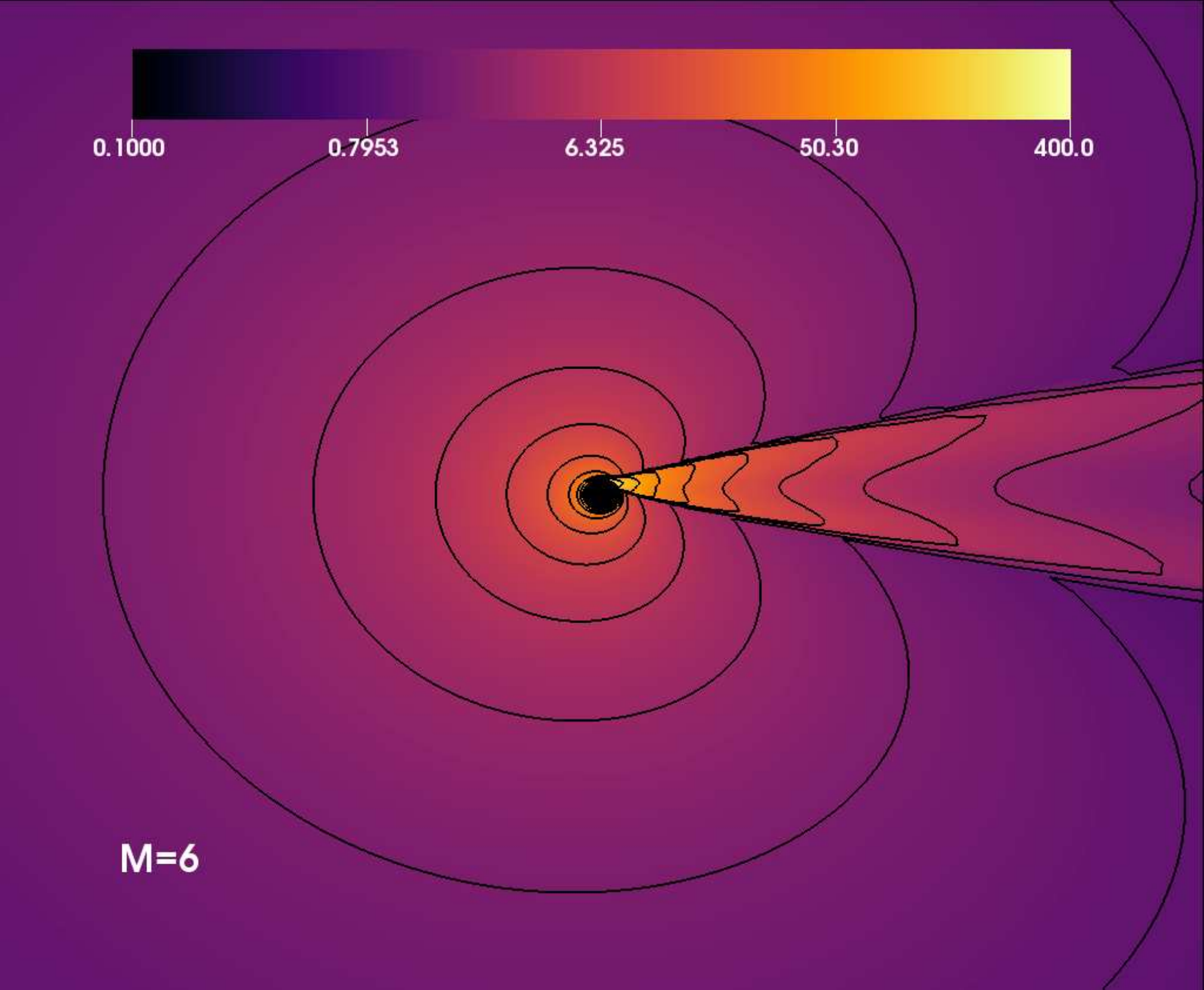,width=6.5cm}
  \caption{BHL accreation from  the subsonic flow ($M_{\infty} < 1$),  sonic flow  ($M_{\infty} = 1$),
    and supersonic flow  ($M_{\infty} > 1$).
    The logarithm of the rest-mass density for  asymptotic Mach number
    $(M_{\infty})$ given in left bottom corner
    of each plot for model $\alpha = -0.5912$ on the equatorial plane ($r - \phi$).
    The opening angle of the shock
    cones decreases with the increasing of $M_{\infty}$ in the supersonic flow.}
%\vspace{1cm}
\label{color1}
\end{figure*}

In order to reveal the effect of asymptotic Mach number $(M_{\infty})$ on the
dynamics of BHL accreation, we need to describe the morphology of the disk
using the initial model $\alpha = -0.5912$ for
GB coupling constant in Table\ref{Inital_Con}. For an adjusted value of  speed of sound $C_{\infty}=0.1$,
there is a critical value of asymptotic Mach number which equals to unity, above and
bellow which shock cones form in the downstream side of computational domain when the
matter is accreated upstream side seen in Fig.\ref{color1}. We plot the logarithm of the rest-mass
density  and linearly spaced isocontours on the equatorial plane.
The disk is initially filled with matter falling from upstream side of the region. It is indicated in
Fig.\ref{color1} that, for  $M_{\infty} = 1$, the shock opening angle
is getting wider and the shock cone is forced to convert into a  bow shock due to ram pressure, gas
pressure and strong gravity. This model indicates that asymptotic Mach number plays a critical role in
the creation of shock cone and its opening angle and it prevents the shock cone created in the upstream region
of the computational domain. Due to the strong gravity, the shock cone becomes attached to the black hole which
produces accretion rate higher in the strong gravitational region \citep{Ruffert1,Ruffert2}.
As clearly shown in Fig.\ref{color1}, higher the asymptotic Mach number
creates a shock cone with a  smaller shock opening angle. The cone with a standing shock 
converts kinetic energy into thermal energy by falling material
toward the rapidly rotating black hole in $4D$ EGB gravity. In addition, the shape of a shock cone very close
to the black hole strongly depends on the black hole spin. As seen in Fig.\ref{color1},
due to the rapidly rotating black hole $a=0.97$, induced frame dragging produces a warping in space-time
as well as the shock cone
connected to the black hole horizon.

%%%%%%%%%%%%%%%%%%%%%%%%%%%%%%%%%%%%%%%%%%%%%%%%%%%%%%%%%%%%%%%%%%%%%%

\begin{figure*}
  \vspace{1cm}
  \center
  \psfig{file=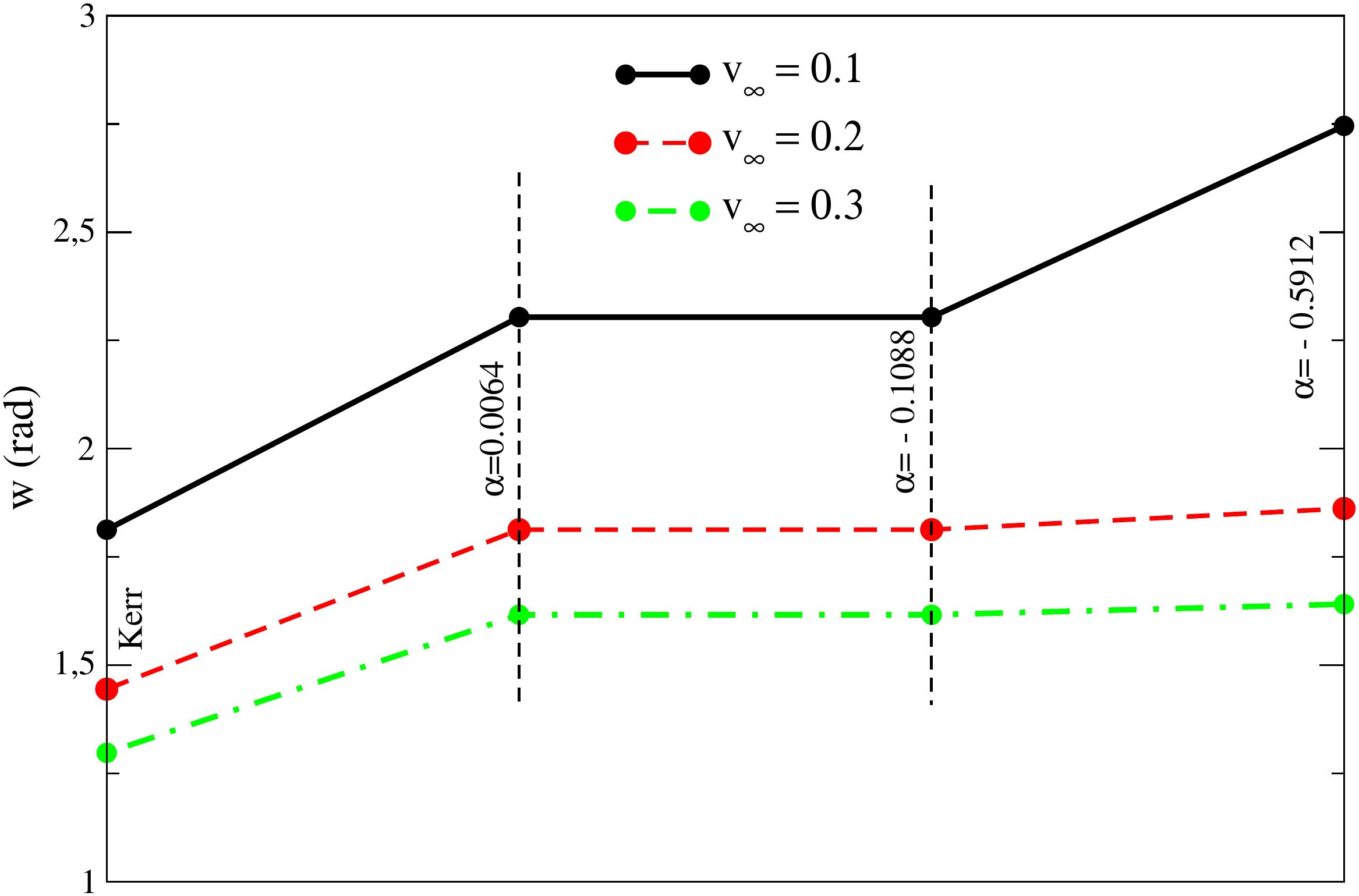,width=10.0cm}
  \caption{Cone width in terms of asymptotic velocity for Kerr and EGB black holes.}
%\vspace{1cm}
\label{cone_with1}
\end{figure*}

In Fig.\ref{cone_with1}, the behavior of the shock cone width with Kerr and $4D$ EGB gravities
for  different values of asymptotic velocities $V_{\infty}$  and the comparison of the results from these two gravities
are given around the rapidly rotating black hole $a=0.97$. It is obvious that $4D$ EGB gravity has a big
influence on the physical properties of the matter around the black hole and this is also confirmed by \citet{Liu1}.
The effect of the asymptotic velocity on the cone width is clearly seen for  the BHL accreation
around the $4D$ EGB gravity as well as the Kerr gravity. Although the comparison of shock opening angle for
different values of $V_{\infty}$ shows the same trend in all models, 
the shock opening angle is smaller around the Kerr black hole.
It is seen that, for the $4D$ EGB gravity, the shock opening angle for 
any value of GB coupling constant $\alpha$
(positive or negative) is greater than the case in Kerr black hole and
as the value of $\alpha$ in negative  direction increases, the deviation from the Kerr geometry increases too.
As it is observed in Fig.\ref{color1}, the shock cone opening angle decreases with increasing of
asymptotic velocity or Mach number seen in Fig.\ref{cone_with1}.

%%%%%%%%%%%%%%%%%%%%%%%%%%%%%%%%%%%%%%%%%%%%%%%%%%%%%%%%%%%%%%%%%%%%%%

\begin{figure*}
  \vspace{1cm}
  \center
  \psfig{file=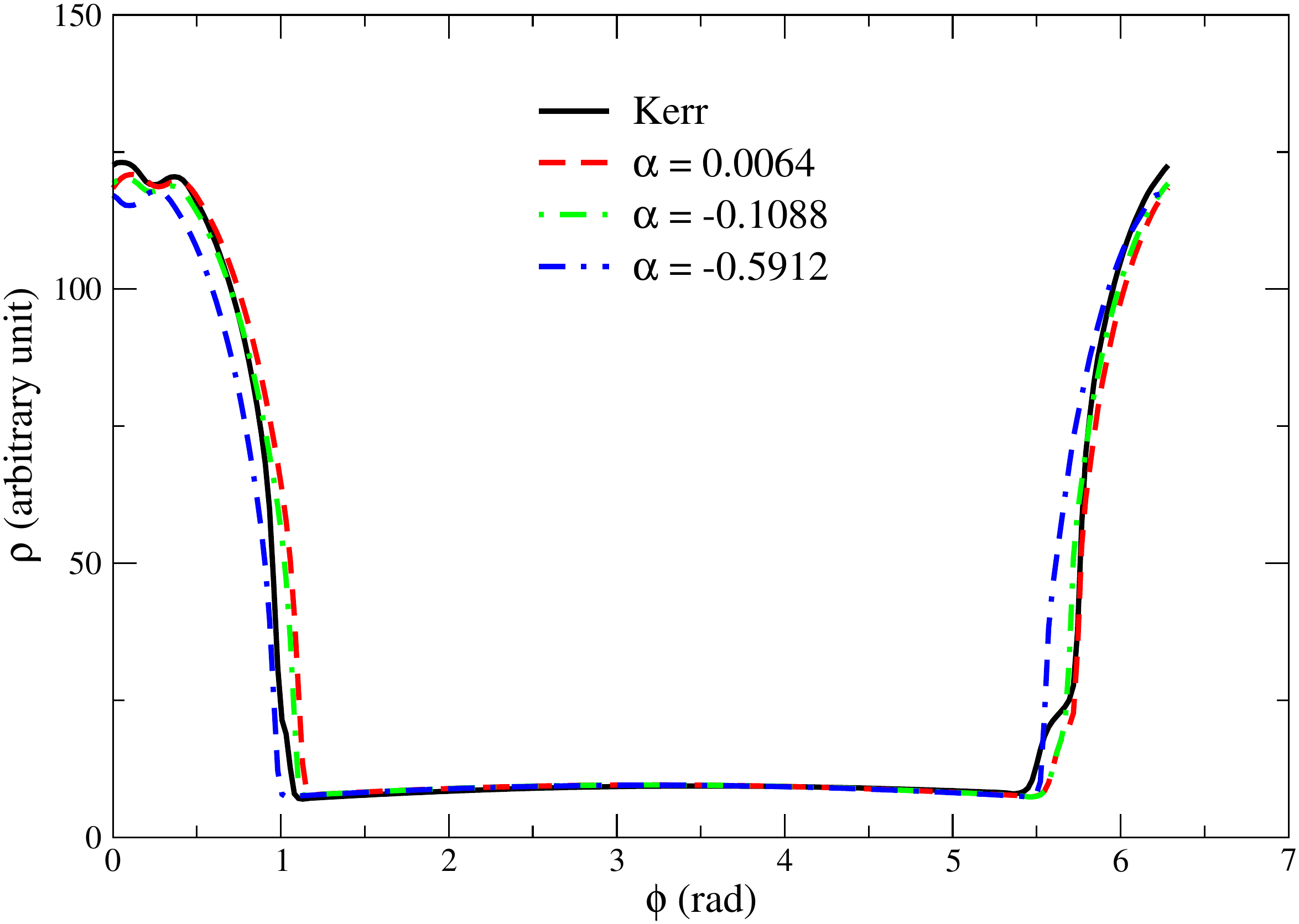,width=10.0cm}
  \caption{The densities of shock cones to showing  the location of discontinuities are plotted
    as a function of angular distance at fixed
    $r=4.78M$ for asymptotic velocity $V_{\infty}=0.1$.}
%\vspace{1cm}
\label{cone_with2}
\end{figure*}

\begin{figure*}
  \vspace{1cm}
  \center
  \psfig{file=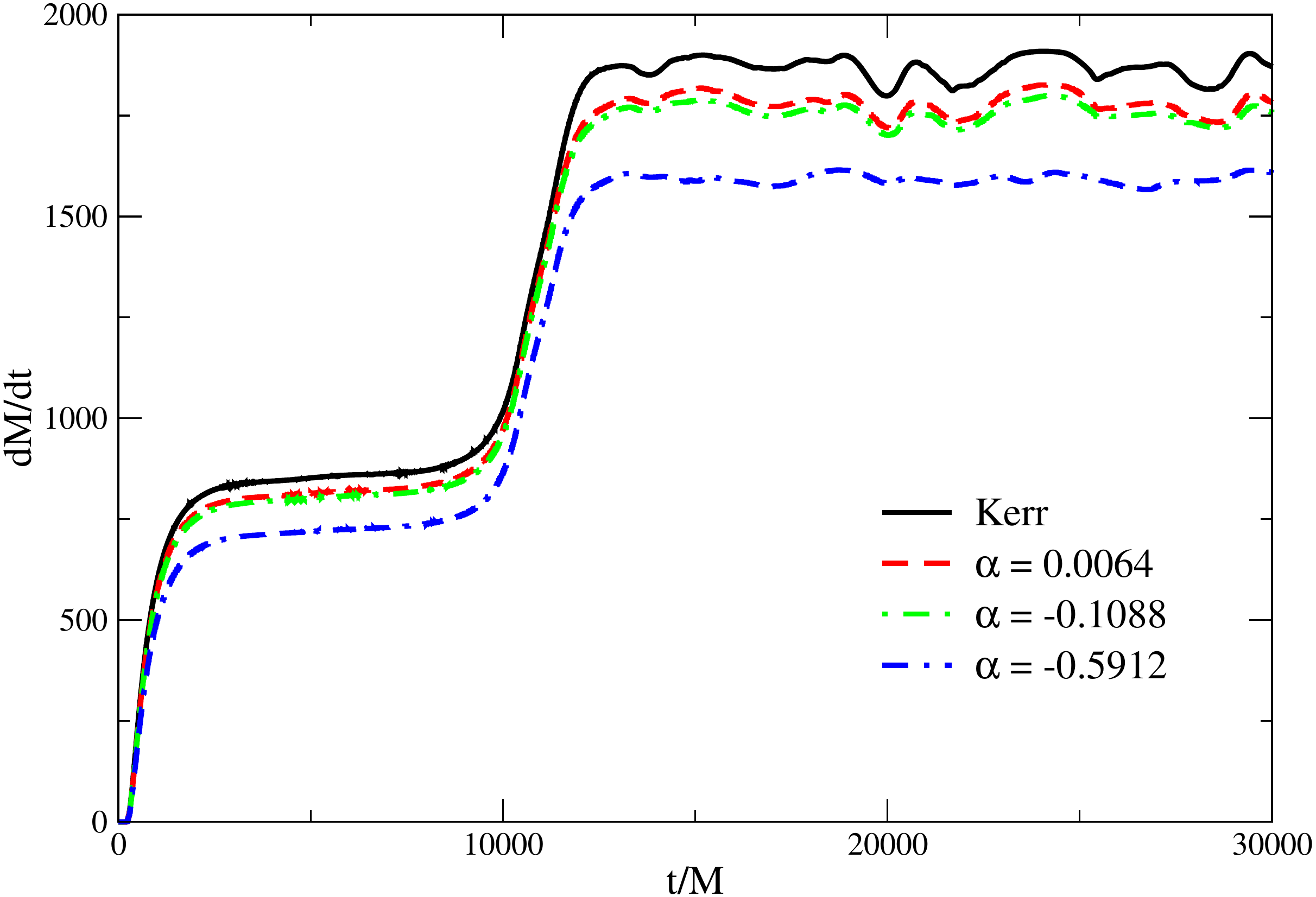,width=10.0cm}
  \caption{Mass accretion rate for asymptotic velocity $V_{\infty}=0.1$.}
%\vspace{1cm}
\label{massAcc1}
\end{figure*}

To reveal the dynamical properties of disk structure and the shock cone location, in Fig.\ref{cone_with2},
we show the rest-mass
density plot at a fixed radial distance $r=4.78M$ for  asymptotic velocity $V_{\infty}=0.1$ at the final
time of the evolution $t=30000M$. As seen in Fig.\ref{massAcc1},
 the shock cone under these conditions formed and reached the steady state around $t=13000M$.
As mentioned earlier, the critical value of asymptotic velocity (Mach number) is $V_{\infty}=0.1$ $(M_{\infty}=1)$.
The critical value of Mach number causes a creation of  strongly oscillating accretion disk in our models.
The effect of the
GB coupling constant  $\alpha$ on the shock cone density at final time is slightly varying as seen in
Fig.\ref{cone_with2}. Kerr black hole case is also plotted on the same figure to compare with the corresponding
results  and find that the rest-mass density is slightly higher in Kerr case. On the other hand,
the angular position of the shock location is shifted for different values of $\alpha$.
The inspection of different curves displayed in Fig.\ref{cone_with2}  shows slight differences in the flow morphology.

%%%%%%%%%%%%%%%%%%%%%%%%%%%%%%%%%%%%%%%%%%%%%%%%%%%%%%%%%%%%%%%%%%%%%%

\begin{figure*}
  \vspace{1cm}
  \center
  \psfig{file=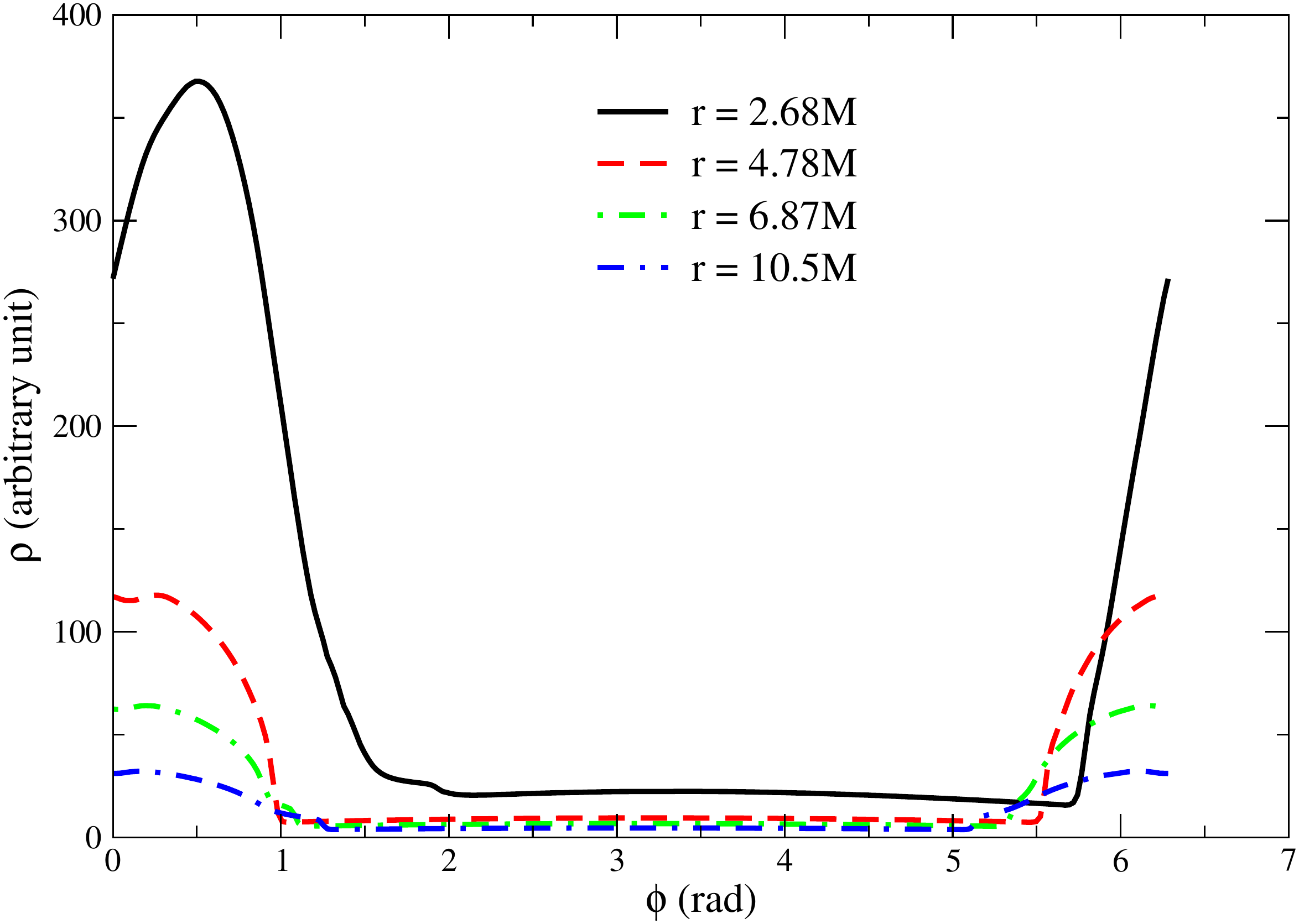,width=10.0cm}
  \caption{The shock cone locations as a function of angular distance for
    asymptotic velocity $V_{\infty}=0.1M$ and Gauss-Bonnet coupling constant
    $\alpha = - 0.5912$ at different $r$.}
%\vspace{1cm}
\label{cone_with3}
\end{figure*}

To complete the overall picture of shock cone structure and morphology of the accretion disk,
Fig.\ref{cone_with3} shows the one-dimensional profile of the rest-mass density at the final time of
simulation at different locations along the radial shells for asymptotic velocity $V_{\infty}=0.1$
and GB coupling constant  $\alpha=- 0.5912$. Clearly,  the strong shock waves are created at the border
of the shock cone. The angular location of the shock cone is slightly changed and
shifted with the radial distance $r$.
As it is expected, the rest-mass density gradually decreases with increasing $r$. The created shock
cone would contribute to the radiation properties of the disk-black hole system.
The shock cone is a very important physical mechanism which converts gravitational energy to the radiation
energy and tapes and excites the oscillation modes. The shock cone could also be used to
explain the erratic spin behavior variety found in different $X-$ray observations \citep{Gergely1, Would1}.

%%%%%%%%%%%%%%%%%%%%%%%%%%%%%%%%%%%%%%%%%%%%%%%%%%%%%%%%%%%%%%%%%%%%%%%%

As seen in the above discussion, the shock cone is the one of the consequences of the BHL accretion
around the black bole in different gravities. The steady-state shock cone forms due to the
effect of gravitation and pressure forces on the equatorial plane. Computing the mass accretion rate
gives us more completed picture about the dynamics and instability of the shock cone around the
rotating black hole. In Fig.\ref{massAcc1}, the time evolution of the mass accretion rate for all the models for
asymptotic velocity $V_{\infty}=0.1$ is shown. The different colors and line styles are used  to separate the
different models. The accretion rate is calculated  using the following expression

\begin{eqnarray}
 \frac{dM}{dt} = -\int_0^{2\pi}\tilde{\alpha}\sqrt{\gamma}\rho u^r d\phi.
\label{GRH7}
\end{eqnarray}

\noindent Although the accretion would lead to an increase in the mass of the black
hole during the evolution, it is assumed that the black hole mass is constant
during the whole simulation. As seen in Fig.\ref{massAcc1}, the shock cone reaches steady-state
around $t=13000M$ in all models and the accreation rate is the highest
in Kerr gravity than EGB gravity for any value of GB coupling constant.
After the cone reaches steady-state, it oscillates around a certain value. The oscillation amplitude is slightly
diminished  in the case of  $\alpha=- 0.5912$ for a fixed value of the black hole rotation parameter $a=0.97$.
The greater the oscillation amplitude inside the shock cone
would lead to a varying $X-$ ray in observed astrophysical phenomena.

%%%%%%%%%%%%%%%%%%%%%%%%%%%%%%%%%%%%%%%%%%%%%%%%%%%%%%%%%%%%%%%%%%%%%%%%

\begin{figure*}
  \vspace{1cm}
  \center
  \psfig{file=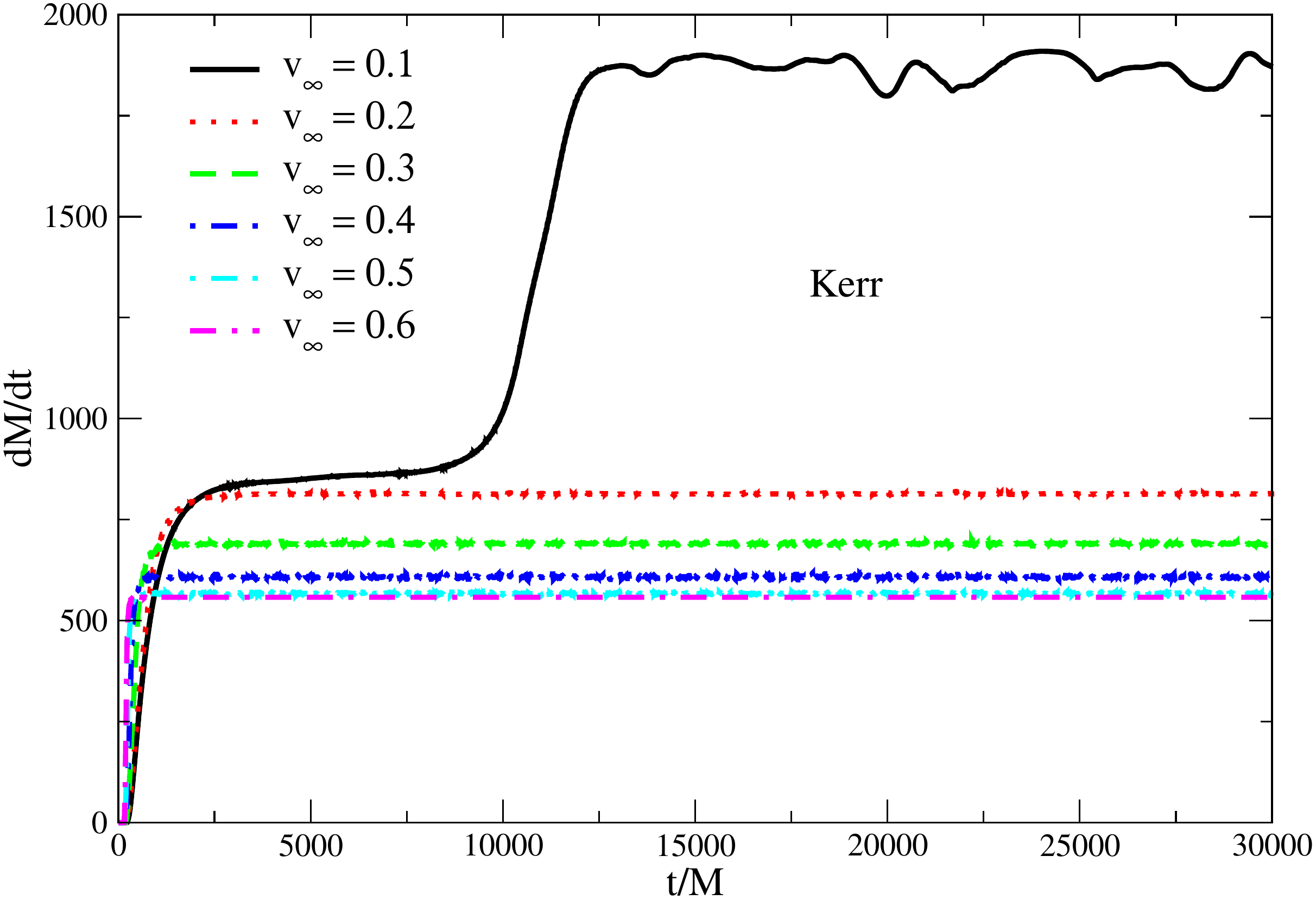,width=7.5cm}  \hspace*{0.15cm}
  \psfig{file=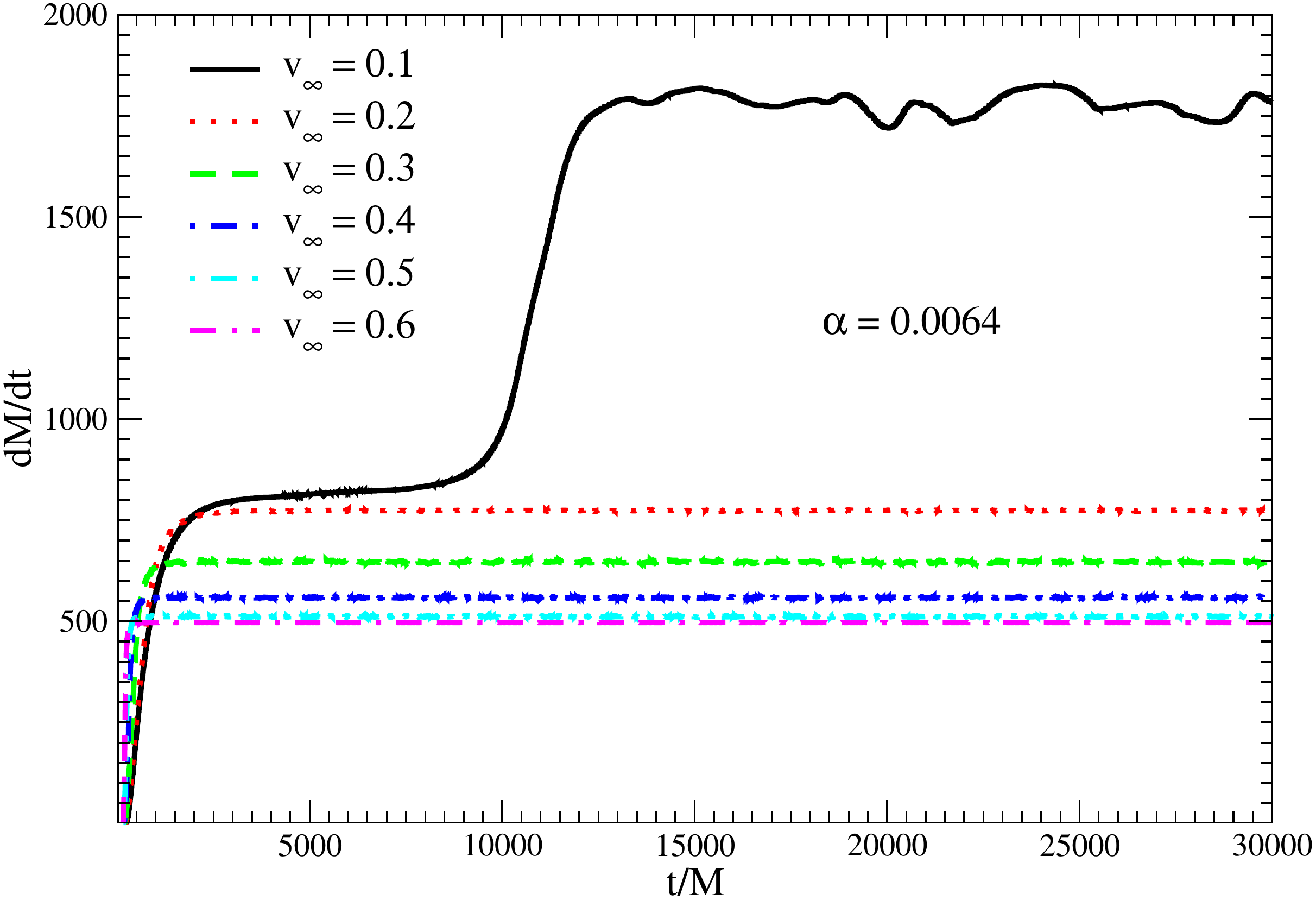,width=7.5cm}\\
  \vspace*{1cm}
  \psfig{file=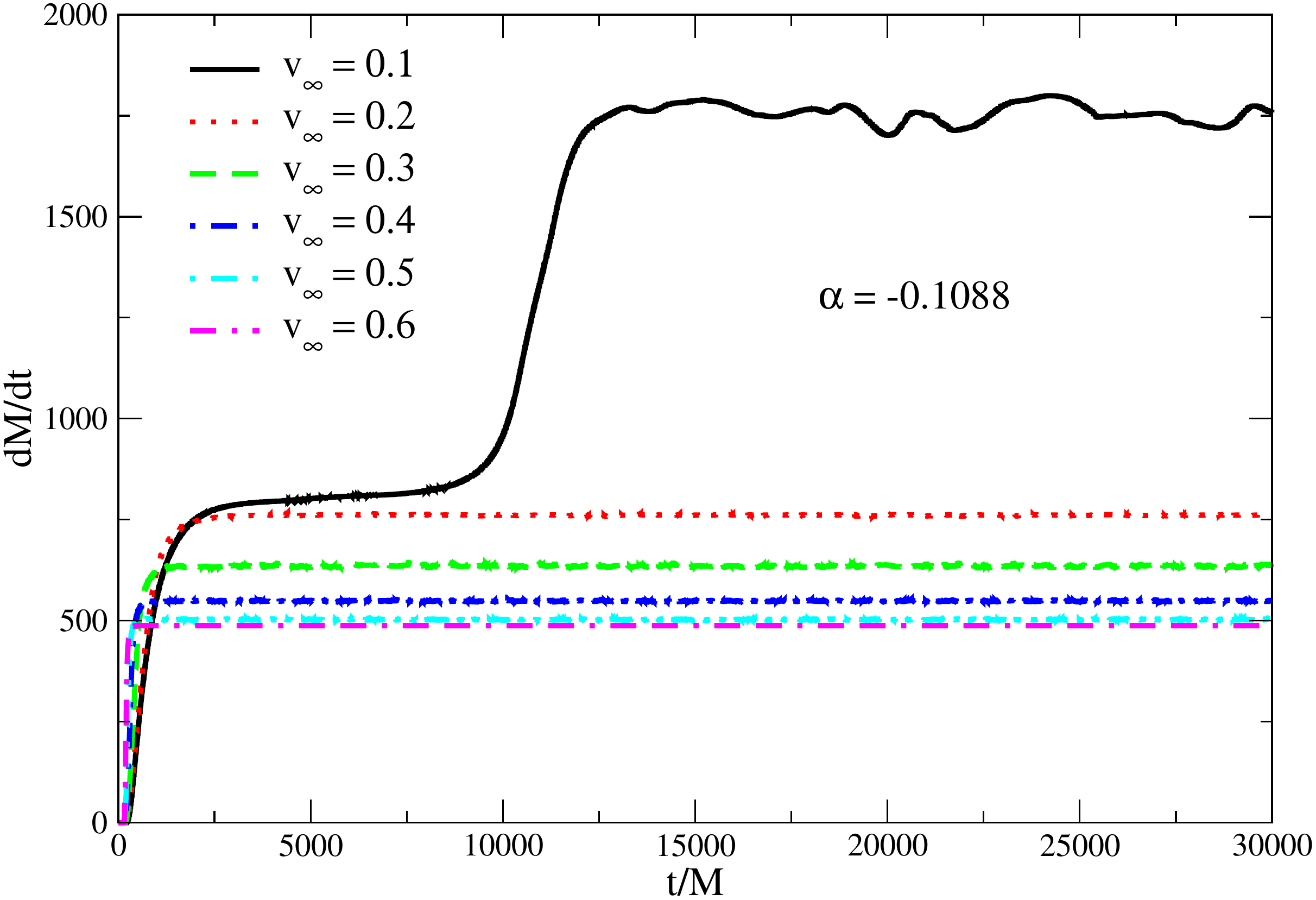,width=7.5cm} \hspace*{0.15cm}
  \psfig{file=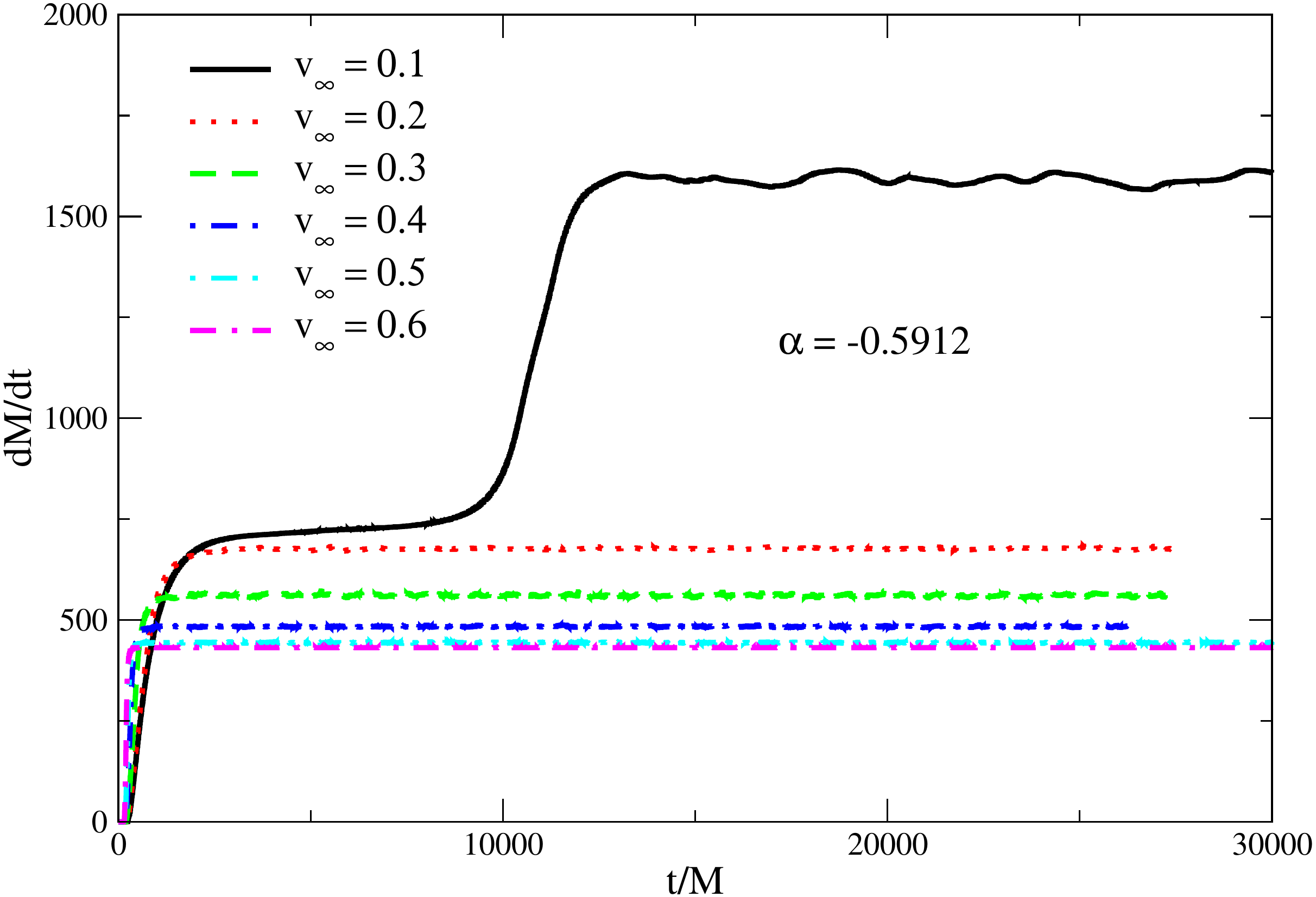,width=7.5cm}
  \caption{Mass accretion rates versus times for Kerr and EGB black holes.}
%\vspace{1cm}
\label{massAcc2}
\end{figure*}

In order to extract more information about the possibility of the oscillation of the shock
cone depending on asymptotic velocity and GB coupling constant we plot Fig.\ref{massAcc2}. Once
the shock cone reaches the steady state, we do not observe any oscillation for all values of $V_{\infty}$
except $V_{\infty}=0.1$. It is also noted that the quantitative
value of accretion rate has the highest value
in Kerr gravity. The accretion rate is getting smaller with increasing the negative value of GB coupling constant.
Another trend, which is clearly seen in Fig.\ref{massAcc2}, is that the steady state is fully developed
around $t=2500M$ for any value of $V_{\infty}$ except $V_{\infty}=0.1$. The stability is first developed around
$t=2500M$ for $V_{\infty}=0.1$ and later the shock cone goes into another unstable region around $t=10000M$. Finally,
the steady state is attained around $t=12500M$. In the steady state, the shock cone
formed in vicinity of the black hole on the equatorial plane is supposed to be in thermodynamical equilibrium.
On the other hand, it is found that the accretion efficiency depends on GB coupling constant $\alpha$.
The black hole with negative $\alpha$ can provide less efficient accreation mechanism which causes the
transformation of the gravitational energy into the electromagnetic energy \citep{Liu1}.

%%%%%%%%%%%%%%%%%%%%%%%%%%%%%%%%%%%%%%%%%%%%%%%%%%%%%%%%%%%%%%%%%%%%%%%%

\begin{figure*}
  \vspace{1cm}
   \center
  \psfig{file=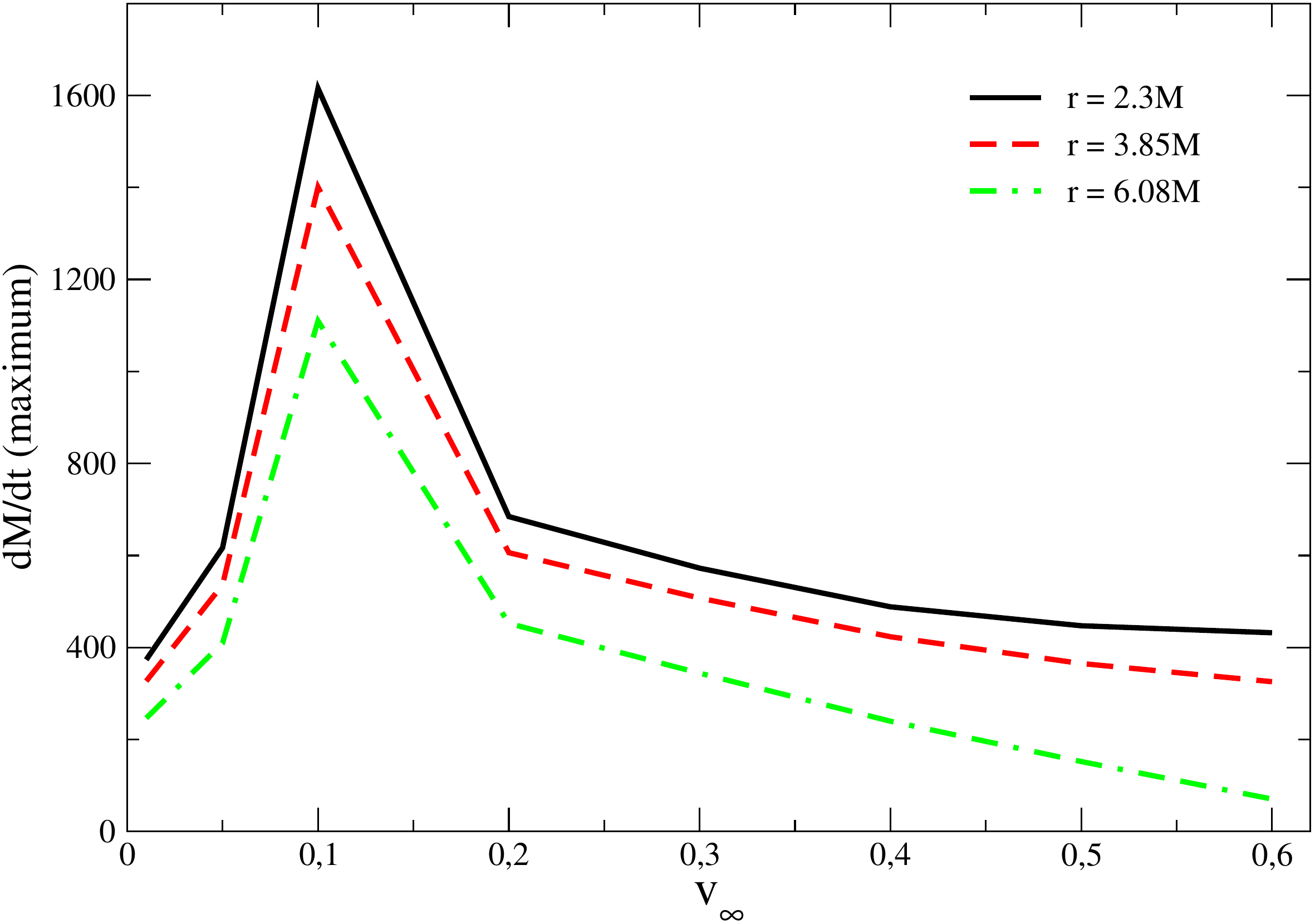,width=10.0cm}
  \caption{The maximum mass accretion rate shown as a function of $V_{\infty}$
    at different radial shells for model $\alpha = -0.5912$.}    
%\vspace{1cm}
\label{massAcc3}
\end{figure*}

Prediction of  the accretion rates from the numerical simulation around the rotating black holes
could be used to define the properties of the black hole as well as $X-$ ray mechanism.
In Fig.\ref{massAcc3}, we plot the behavior of the accretion rate with an asymptotic velocity at
different locations along the radial distance on the disk.
As seen in the Fig.\ref{massAcc3}, the accretion rates exponentially decrease around
the sonic point which occurs at $V_{\infty}=0.1$ .
The reason of this decrease is due to the gravitational force and the pressure
forces that they cannot balance each other and it causes a sudden change in the flow morphology
when the velocity goes in the direction of subsonic or supersonic region.  The gravitational
force is dominant in the subsonic region while the gas pressure is dominant in the supersonic region,
especially far away from the strong gravitational region.

%%%%%%%%%%%%%%%%%%%%%%%%%%%%%%%%%%%%%%%%%%%%%%%%%%%%%%%%%%%%%%%%%%%%%%%%

\begin{figure*}
  \vspace{1cm}
  \center
  \psfig{file=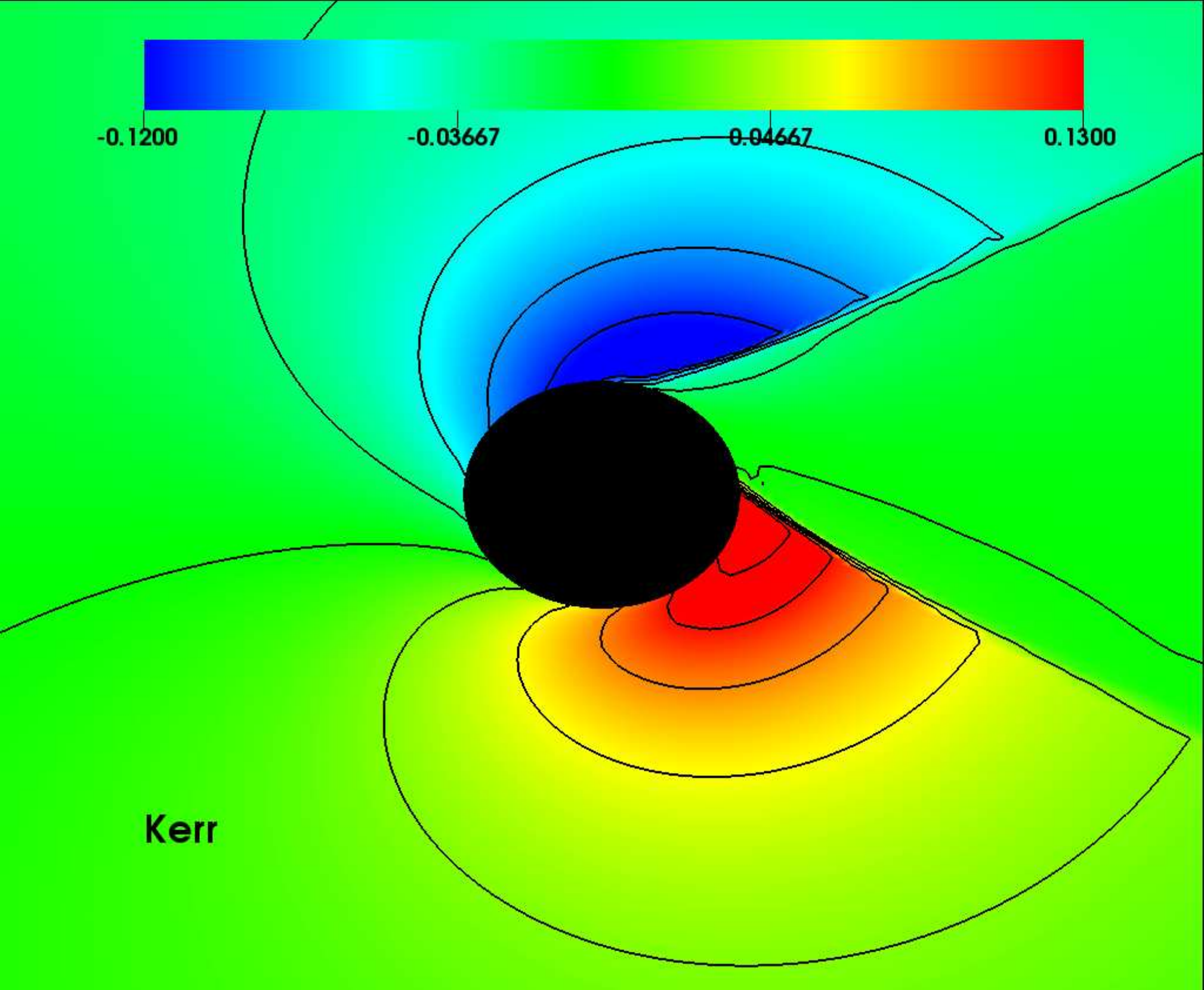,width=7.5cm}
  \psfig{file=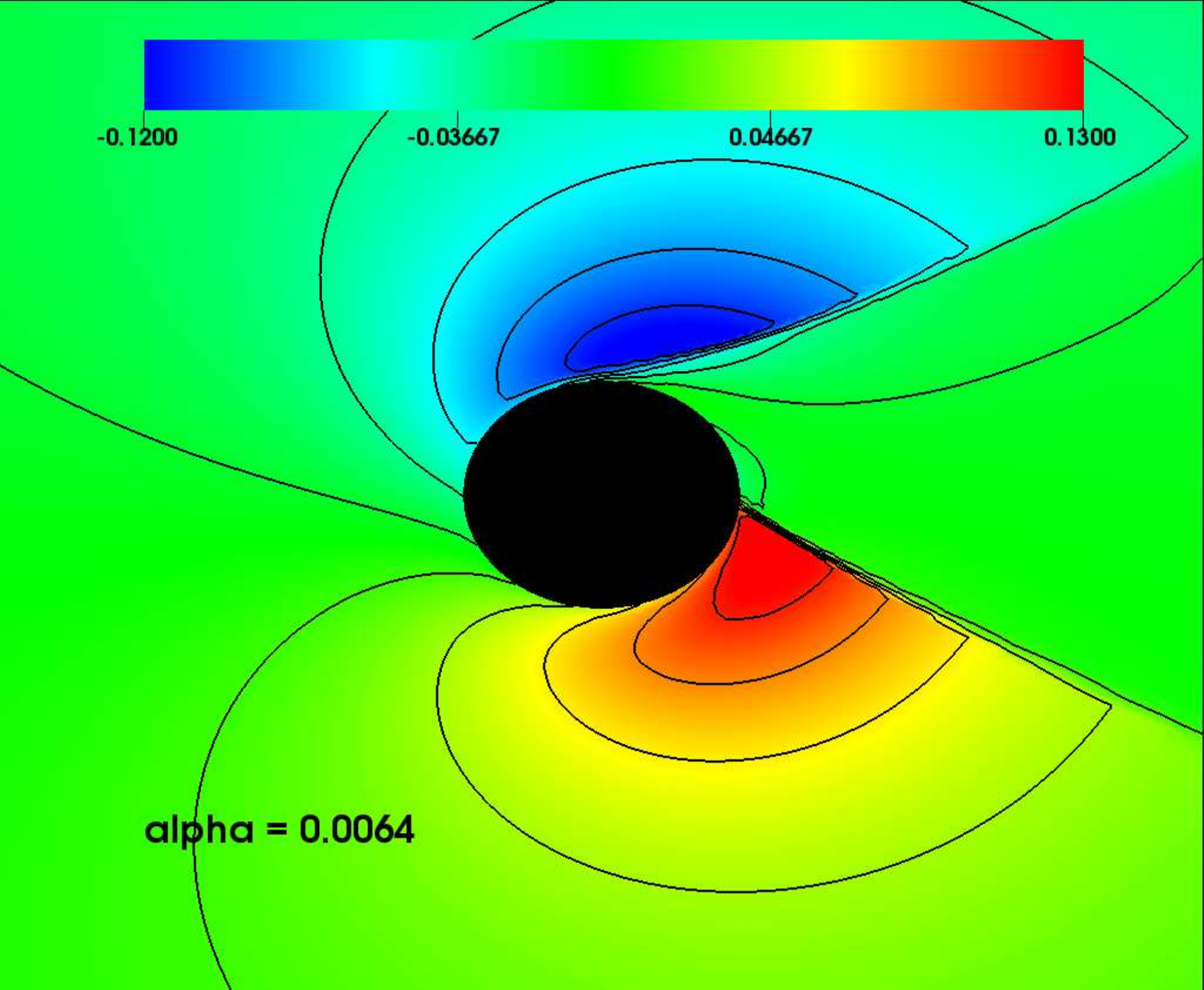,width=7.5cm}
  \psfig{file=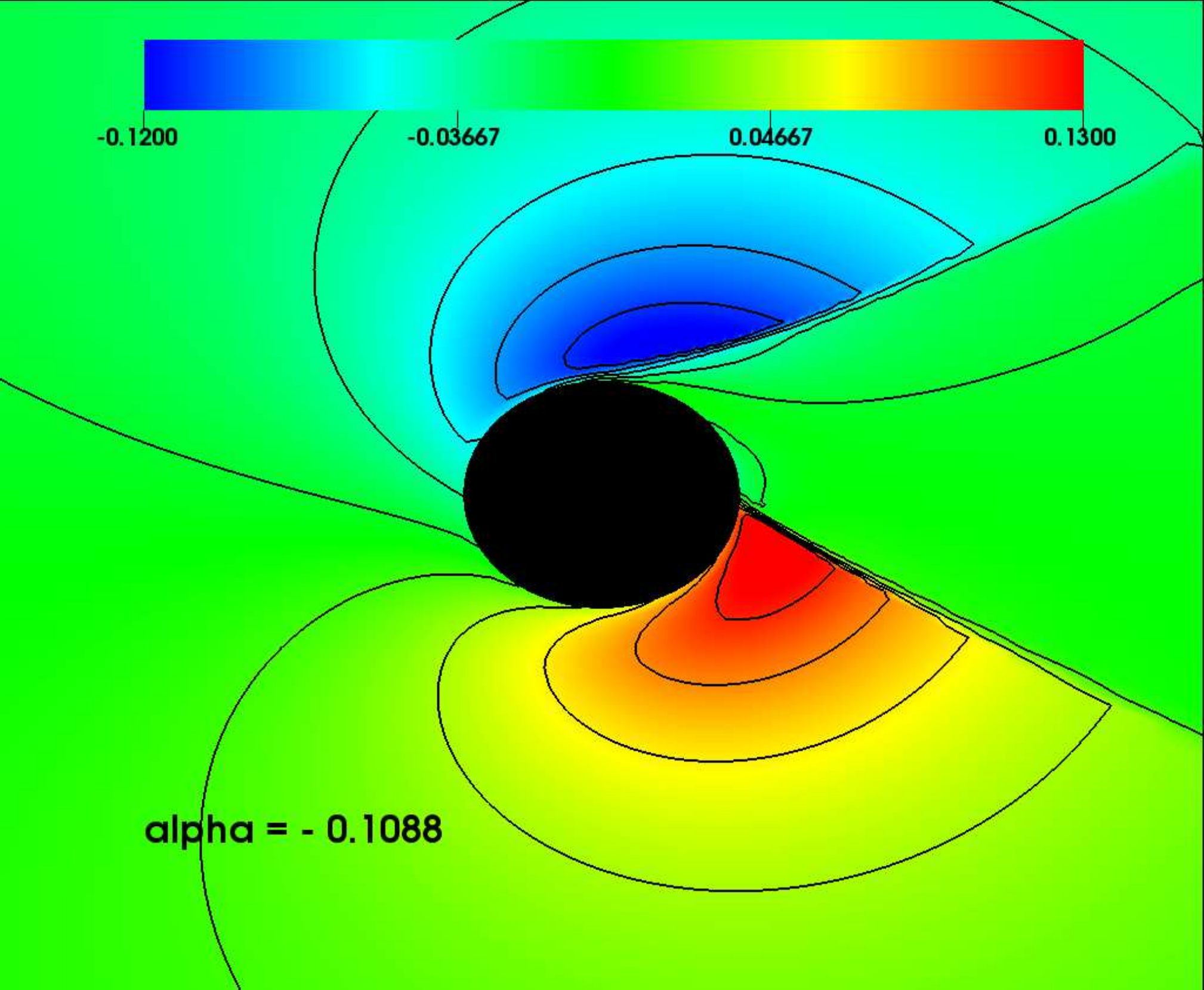,width=7.5cm}
  \psfig{file=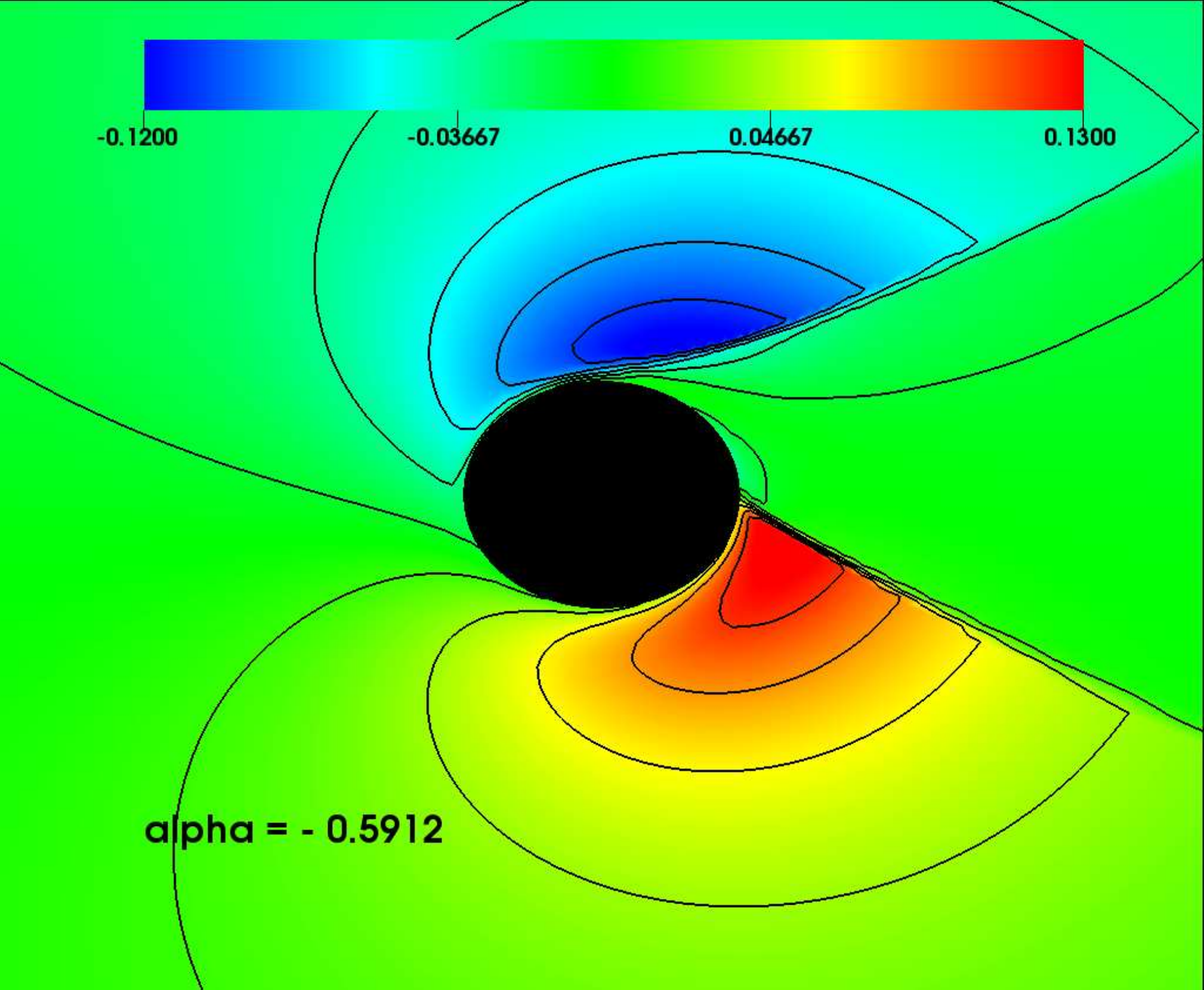,width=7.5cm}  
  \caption{ The radial velocity profiles for Kerr and EGB black holes with $V_{\infty} = 0.2$
    on the equatorial plane ($r - \phi$) a long time later after the disk reached to the steady state.}
%\vspace{1cm}
\label{color2}
\end{figure*}

The radial velocity profile is an important indicator to extract the violent features of the accreation mechanism
and shock cone created around the black hole in different gravities. The kinetic energy of the violent
motion depicts the thermal efficiency of the system. The color and counter plots of the
radial velocity profile of the shock cone in strong gravitational region are shown in Fig.\ref{color2} for
Kerr and EGB gravities. The shock cone structure and radial velocity can be seen in the figure. 
While the matter is falling toward the black hole in one side of the shock cone location
(positive velocity)
it moves away from the other location (negative velocity).
But  the matter inside the shock cone oscillates in steady state. The dynamically stable flow
seen in Fig.\ref{color2} produces a continuum mechanism  for creation of $X-$ ray around
the rapidly rotating black holes. It is also shown that the behavior of the counter lines inside the shock cone
around the Kerr black hole is slightly different than the ones in EGB gravity.
In addition, there is no noticeable effect observed
for the different values of GB coupling constant $\alpha$ in the strong gravitational region.

%%%%%%%%%%%%%%%%%%%%%%%%%%%%%%%%%%%%%%%%%%%%%%%%%%%%%%%%%%%%%%%%%%%%%%%

\begin{figure*}
  \vspace{1cm}
  \center
  \psfig{file=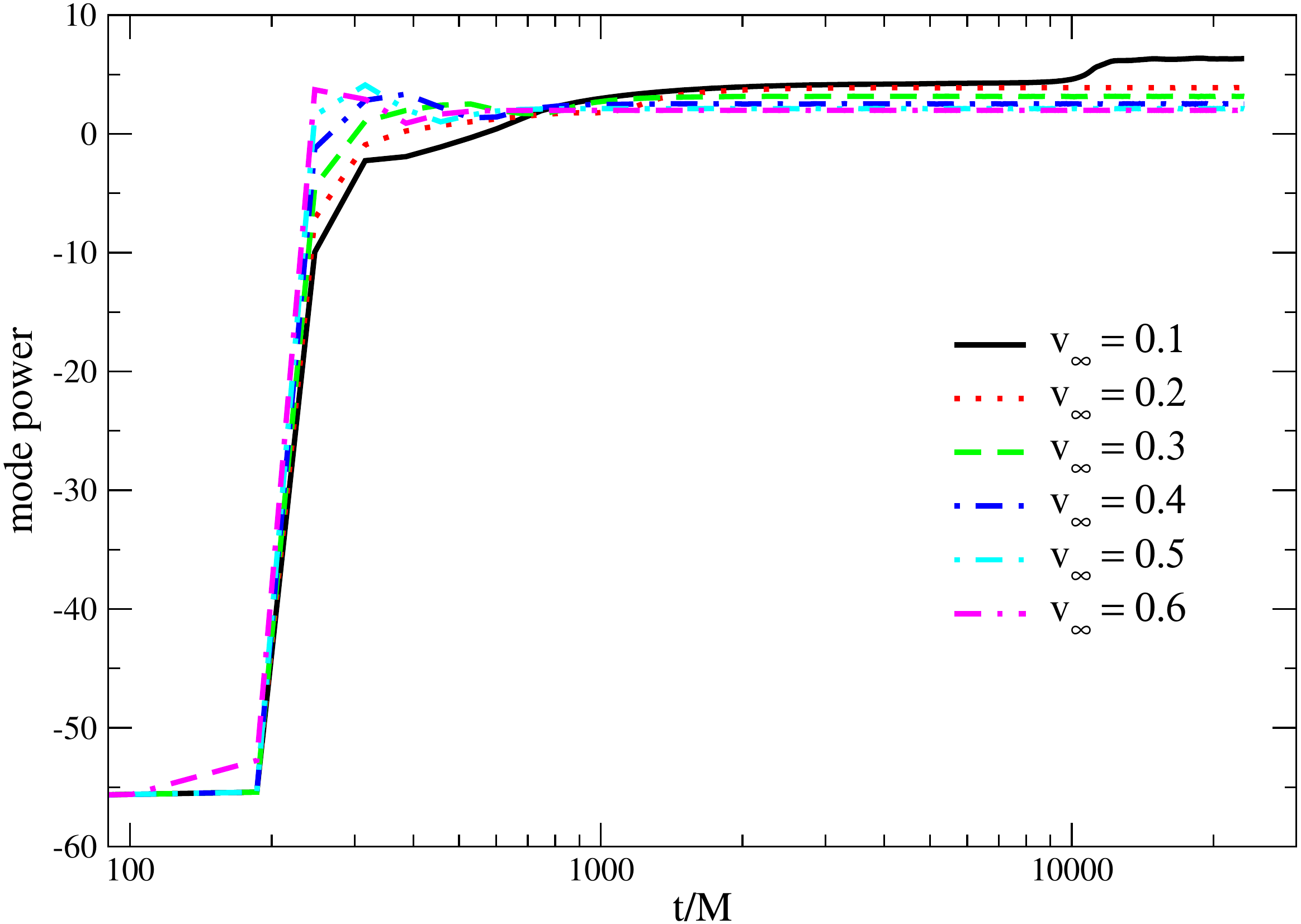,width=10.0cm}
  \caption{ The evolution of $m=1$ mode power for EGB black hole with $\alpha=-0.5912$.}
%\vspace{1cm}
\label{mode_power1}
\end{figure*}

\begin{figure*}
  \vspace{1cm}
  \center
  \psfig{file=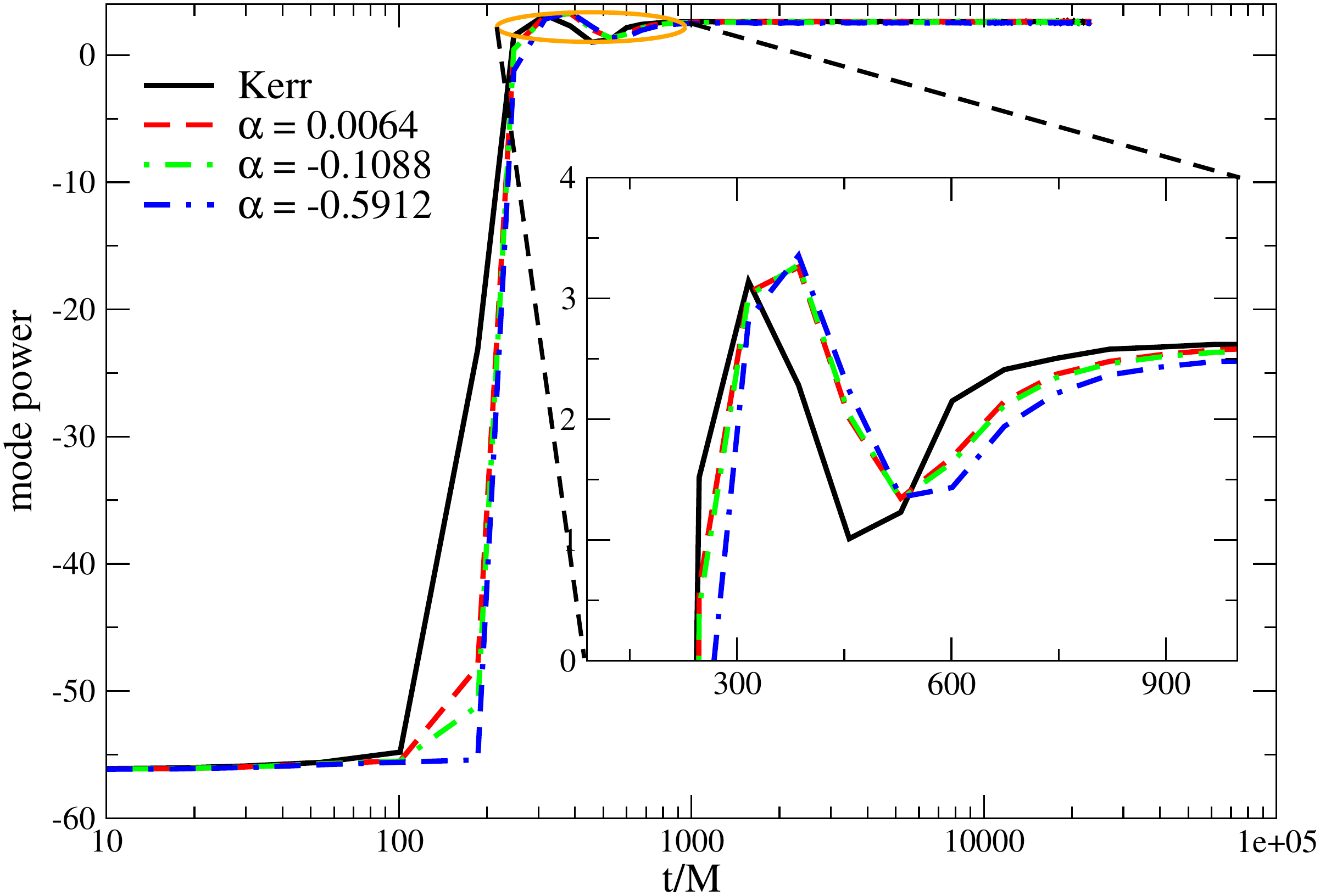,width=10.0cm}
  \caption{ The evolution of $m=1$ mode power for Kerr and EGB black holes with $V_{\infty}=0.4$.}
%\vspace{1cm}
\label{mode_power2}
\end{figure*}

The shock cone instabilities can be calculated by performing  a power mode analysis on the steady state shock cone
around the black hole in Kerr and EGB gravities. This can be done by defining the azimuthal wavenumber
 and finding the saturation points in the oscillating system \citep{Donmez4}. The growth of the instability
modes for the azimuthal wavenumber $m=1$  in case of the BHL accretion in Kerr and EGB gravities is given
in Figs.\ref{mode_power1} and \ref{mode_power2}. As expected, the mode rapidly increases due to BHL accretion onto
the black hole. The mode and saturation point are developed for  all the values of asymptotic velocity about
$t=250M$ except $V_{\infty}=0.1$ as seen in Fig.\ref{mode_power1}. The power mode for asymptotic velocity
$V_{\infty}=0.1$ keeps growing until $t=13000M$ and  it reaches saturation. The shock cone and the disk do not show any
remarkable oscillation after the system reaches a saturation in all simulations. As seen in Fig.\ref{mode_power2},
the power mode analysis of the shock cone in EGB gravity shows slightly different behavior than Kerr gravity.
The instability is developed at a later time in the EGB gravity and it saturates slightly later than the Kerr gravity
(cf. the inset of Fig.\ref{mode_power2}).
In addition, we did not observed any substantial difference in the behavior  of the power mode analysis for
different $\alpha$. Fig.\ref{mode_power2} also shows that the instability is developed at a slightly later time than
the other ones for $\alpha=-0.5912$.

%%%%%%%%%%%%%%%%%%%%%%%%%%%%%%%%%%%%%%%%%%%%%%%%%%%%%%%%%%%%%%%%%%%%%%%
%%%%%%%%%%%%%%%%%%%%%%%%%%%%%%%%%%%%%%%%%%%%%%%%%%%%%%%%%%%%%%%%%%%%%%%

\section{Astrophysical Motivation}
\label{AstroMot}

There are possible candidates  in which our numerical simulation results could be used to identify properties of disk
structure and radiation properties.
Firstly, the variable $X-$ ray sources are observed in nearby galaxies \citep{Walton1}. They are ultra-luminous
$X-$ ray sources and contain either
intermediate black holes ($M_{BH} \sim 10^2-10^5 M_{\odot}$) or massive stellar black holes
($M_{BH} \sim 30-100M_{\odot}$). NGC $1313$ $X-1$ could be a quite massive ($M_{BH} \sim 70-100M_{\odot}$),
accreating close to the Eddington limit and contains a hot inner disk \citep{Bachetti1}. The
rotation parameter of NGC $1313$ $X-1$ is extreme and defined as $0.93 \leq a \leq 0.96$ \citep{Haydari2}.
Secondly, MAXI $J1803-298$ is a newly discovered rapidly rotating black hole candidate with a spin parameter of 
$0.991$ \citep{Feng2}.
Lastly, many of the super-massive black holes examined seem to be rapidly rotating  ones \citep{Reynolds1}.
The black hole spin parameter used in our numerical simulation for Kerr and 4D EGB black holes is $0.97$
which falls in the above observed values.
The spin is an important parameter constraining the model properties and gives a new insight about
accreation dynamics. 
Our numerical solution is also interesting to  use
the exploring the features of electromagnetic
radiation and to constrain the black hole parameters depending on GB coupling constant $\alpha$.
For example, it is possible to measure the deviation of the angular momentum of the massive stellar
Kerr black hole using the continuum fitting \citep{Bambi1}. The observational constrain of EGB gravity was mainly
discussed in \citet{Feng1,Timothy1} for various physical systems.

%%%%%%%%%%%%%%%%%%%%%%%%%%%%%%%%%%%%%%%%%%%%%%%%%%%%%%%%%%%%%%%%%%%%%%%
%%%%%%%%%%%%%%%%%%%%%%%%%%%%%%%%%%%%%%%%%%%%%%%%%%%%%%%%%%%%%%%%%%%%%%%

\section{Discussion and Conclusion}
\label{Conclusion}
%%%%%%%%%%%%%%%%%%%%%%%%%%%%%%%%%%%%%%%%%%%%%%%%%%%%%%%%%%%%%%%%%%%%%%%
In this paper, we investigate  and compare BHL accreation onto the Kerr and EGB
rotation black holes with the rotation
parameter $a=0.97$ in  a strong gravitational region using the non-axisymmetric
hydrodynamical simulation. We use the
perfect fluid equation of state with adiabatic index $\Gamma = 4/3$. We study the
effect of different values of 
GB coupling constants in  subsonic, sonic, and supersonic flow regions
to reveal the accretion disk dynamics and shock cone structure around the
rapidly rotating black holes. 

To reveal the shock cone structure, physical properties, and the radiation properties we have
studied the mass accretion rate, the behavior of the rest-mass density, the shock
opening angle, radial velocity of the flow, and power mode for different values of
GB coupling constants and asymptotic velocities with a fixed value of the rotation parameter.
The effect of the GB coupling constant $\alpha$ on the shock cone properties is discussed and
compared with the Kerr black hole in general relativity. It is found that the shock opening angle
is greater in EGB gravity for any value of $\alpha$ when we compare it with Kerr gravity.
Besides, the shock opening angle decreases with increasing  asymptotic velocity of Mach number in both Kerr and
EGB gravities. On the other hand the accretion rates are nearly constant  in all models after the steady state
is reached and the amount of the rate is getting less for the greater value of $\alpha$ in negative direction.
The accreation rate is the higher around the Kerr black hole than EGB black hole for
any value of GB coupling constant. After the shock cone reaches to steady-state, it oscillates
around a certain value. The oscillation amplitude is slightly diminished for  $\alpha=- 0.5912$.
The high the oscillation amplitude in shock cone could be a good candidate to observe a time varying
$X-$ rays  in astrophysical phenomena.

The shock location of the cone which is attached  to the black hole horizon
is a natural physical system which transports the angular momentum of the accreating
gas emerging close to the black hole. Therefore the transportation is responsible for the mass accreation
onto the black hole. We find the efficiency of the accretion rate to be a strong function of the
asymptotic velocity (Mach number). Even though the accretion rate exponentially decreases in the subsonic and
supersonic regions, it reaches maximum value  at sonic location which occurs at $V_{\infty}=0.1$.
The gravitational force has a big influence  in the subsonic region while the gas pressure is 
dominant in the supersonic region, especially far away from the strong gravitational region.

Finally, it is shown in our numerical simulations that the axisymmetric accretion disk with the shock cone
around the rotating black hole in Kerr and EGB gravities generates more violent phenomena and it gets hotter
at small value of $\alpha$ than the value of $\alpha=- 0.5912$.
The more violent model could be a good candidate to explore
the physical features of the electromagnetic radiation of NGC $1313$ $X-1$ and $X-2$.
The black hole spin parameter used in our numerical simulation for Kerr and 4D EGB black holes is $0.97$
which falls in the observed value of NGC $1313$ $X-1$ and $X-2$ and MAXI $J1803-298$.

%%%%%%%%%%%%%%%%%%%%%%%%%%%%%%%%%%%%%%%%%%%%%%%%%%%%%%%%%%%%%%%%%%%%%%%%%%%%%%%%%%%%%%%%
%%%%%%%%%%%%%%%%%%%%%%%%%%%%%%%%%%%%%%%%%%%%%%%%%%%%%%%%%%%%%%%%%%%%%%%%%%%%%%%%%%%%%%%

\section*{Acknowledgments}
All simulations were performed using the Phoenix  High
Performance Computing facility at the American University of the Middle East
(AUM), Kuwait.\\

\bibliography{paper.bib}

\end{document}